%% file: word2vec-based-approx.tex
\documentclass[conference]{IEEEtran}
\IEEEoverridecommandlockouts

\usepackage{xspace}
\usepackage{xifthen}
\usepackage{subfig}
\usepackage{listings}
\usepackage{subfig}
\usepackage{url}
\usepackage{xcolor}
\usepackage{booktabs} 
\usepackage{mathtext}
\usepackage{graphicx}
\usepackage{array}
\usepackage{url}
\usepackage{amsmath}
\usepackage{cite}

\newcommand{\myparagraph}[2][]{\ifthenelse{\isempty{#1}}{\vspace{1ex} \noindent {\bf #2}}{\noindent {\bf #2}}}
\usepackage{amsfonts}

\newcommand{\futurenote}[1]{
}

\newcommand{\reconsider}[1]{
}
\newcommand{\thunote}[1]{{\textcolor{red}{[Thu: #1]}}}
\newcommand{\guangyannote}[1]{{\textcolor{blue}{[Guangyan: #1]}}}
\newcommand{\yongfengnote}[1]{{\textcolor{cyan}{[Yongfeng: #1]}}}

\newcommand*{\affaddr}[1]{#1} 

\def\BibTeX{{\rm B\kern-.05em{\sc i\kern-.025em b}\kern-.08em
		T\kern-.1667em\lower.7ex\hbox{E}\kern-.125emX}}

\begin{document}

\title{Similarity Driven Approximation for Text Analytics}

\author{
    \IEEEauthorblockN{Guangyan Hu}
	\IEEEauthorblockA{\textit{Rutgers University} \\
		New Brunswick, NJ \\
		gh279@cs.rutgers.edu}
	\and
	\IEEEauthorblockN{Yongfeng Zhang}
	\IEEEauthorblockA{\textit{Rutgers University} \\
		New Brunswick, NJ \\
		yz804@cs.rutgers.edu}
	\and
	\IEEEauthorblockN{Sandro Rigo}
	\IEEEauthorblockA{\textit{University of Campinas} \\
		Campinas - SP, Brazil \\
		srigo@unicamp.br}
	\and
	\IEEEauthorblockN{Thu D. Nguyen}
	\IEEEauthorblockA{\textit{Rutgers University} \\
		New Brunswick, NJ \\
		tdnguyen@cs.rutgers.edu}
}

\maketitle

\input{abstract}

\input{intro}
\input{related}
\input{technical}

\input{tasks}

\input{discussions}
\input{implementation}

\input{evaluation}
\input{conclusion}

\bibliographystyle{IEEEtran}
\bibliography{word2vec-based-approx}

\end{document}

%% file: abstract.tex
\begin{abstract}
Enterprises are increasingly seeking to extract insights for decision making from text data sets. Yet, processing large text data sets using sophisticated algorithms is computationally expensive.
In this paper, we propose and evaluate a framework called EmApprox that uses approximation to speed up the processing of a wide range of queries over large text data sets. The key insight is that different types of queries can be approximated by processing subsets of data that are most similar to the queries. EmApprox builds a general index for a data set by learning a natural language processing model, producing a set of highly compressed vectors representing words and subcollections of documents. Then, at query processing time, EmApprox uses the index to guide sampling of the data set, with the probability of selecting each subcollection of documents being proportional to its {\em similarity} to the query as computed using the vector representations. We have implemented a prototype of EmApprox as an extension of the Apache Spark system, and used it to approximate three types of queries: aggregation, information retrieval, and recommendation. Experimental results show that EmApprox's similarity-guided sampling achieves much better accuracy than random sampling. Further, EmApprox can achieve significant speedups if users can tolerate small amounts of inaccuracies. For example, when sampling at 10\%, EmApprox speeds up a set of queries counting phrase occurrences by almost 10x while achieving estimated relative errors of less than 22\% for 90\% of the queries.
\end{abstract}

%% file: intro.tex
\section{Introduction}
\label{sec:introduction}

\myparagraph{Motivation.} Enterprises are increasingly seeking to extract insights for decision making from text data sets. At the same time, data is being generated at an unprecedented rate, so that text data sets can get very large. For example, the Google Books Ngram data set contains 2.2~TB of data~\cite{GNgram} and the Common Crawl corpus contains petabytes of data~\cite{CommonCrawl}. Processing such large text data sets using sophisticated algorithms is computationally expensive.

The above challenge is exacerbated when it is desirable to run different types of queries against a data set, making it expensive to build multiple indices to speedup query processing. For example, given a data set comprising user reviews on products, an enterprise may want to count positive vs.~negative reviews, use the reviews to make recommendations, or retrieve reviews relating to a particular product~\cite{McAuley:2013}. Currently, a different index is required for quickly answering each of these query types.

\myparagraph{EmApprox.} In this paper, we propose a framework called EmApprox to speed up a wide range of queries over large text data sets. The key idea behind EmApprox is to build a general index that guides the processing of a query toward a subset of the data that is most {\em similar} to the query. For example, consider a query that seeks to count the number of occurrences of a given phrase. EmApprox would select a sample of the data set, preferentially choosing items most similar to the query phrase, count the occurrences in the sample, and use the count to estimate the number of occurrences in the entire data set. Clearly, the result is {\em approximate} so that users of EmApprox would need to tolerate some imprecision in the estimated results. EmApprox allows users to trade off precision and performance by adjusting the sampling rate.

Our approach is related to the many approximate query processing (AQP) systems that answer aggregation queries over relational data sets by processing estimators with error bounds using samples of the data~\cite{BlinkDB, Sapprox, ApproxHadoop, peng2018aqp++}. In essence, one can think of EmApprox as extending AQP to text analytics. EmApprox supports the estimation of errors bounds when possible; e.g., for aggregation queries. However, EmApprox can also be used in scenarios where it is not possible to estimate error bounds such as information retrieval, making it widely applicable to many different text analytic queries/applications.

\begin{figure}[!tbp]
	\hspace*{.8cm}  
	\includegraphics[width=7cm]{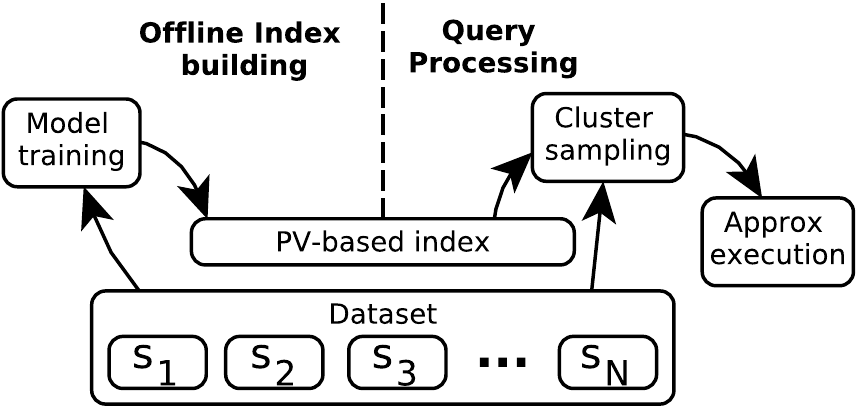}
	\caption{Overview. {${S_n}$} are subcollections of documents.}
	\label{fig:overview}
	\vspace{-.5cm}
\end{figure}

\myparagraph{System overview.} Figure~\ref{fig:overview} gives an overview of EmApprox. As mentioned above, EmApprox executes a query on a sample of the data set to reduce query processing time. Straightforward use of random sampling can lead to large errors, however, when sampling from a skewed distribution~\cite{Sapprox}. To mitigate this issue, EmApprox builds an index {\em offline}, then consults the index at query processing time to guide sampling toward subsets of data that are most similar to the query.

Specifically, EmApprox uses a natural language processing (NLP) model~\cite{doc2vec} to learn vector representations for unique words and documents. The resulting vectors can be composed and used to compute a similarity metric. Then, assuming that the data set is partitioned into a number of subcollections as shown in Figure~\ref{fig:overview}, EmApprox computes a vector for each subcollection from the vectors of the documents contained in it.\footnote{Data sets may be partitioned into subcollections for a variety of reasons, including storage in a distributed file system such as HDFS~\cite{hdfs}.}
The final index contains vectors for unique words together with vectors for the subcollections.

At query processing time, EmApprox computes a vector for the query using vectors of the words in the query. It then computes a sampling probability for each subcollection that is proportional to the similarity between the subcollection and the query using their vector representations. Finally, it selects a sample of subcollections using their sampling probabilities.


EmApprox uses locality-sensitive hashing~(LSH) to hash each real-valued vector to a bit vector~\cite{ANNLsh} to reduce the storage overhead of the index. LSH works well because it preserves the distance between the original vectors. Computing similarity using LSH bit vectors is also extremely cheap; it is simply the Hamming distance of two bit vectors that can be computed efficiently using {\tt XOR}. This optimization has greatly increased the scalability and efficiency of EmApprox's index.

\myparagraph{Queries.} We have implemented a prototype of EmApprox and used it to support approximate processing for three different types of queries: (1) aggregation queries that count occurrences within a text data set, (2) retrieval queries, both Boolean and ranked, that retrieve relevant documents, and (3) recommendation queries that predict users' ratings for products. For aggregation queries, we show how to compute estimated error bounds along with the approximate results. We also show that the training objective of PV-DBOW~\cite{doc2vec}, the specific NLP model that we use, is directly related to minimizing the variance of the estimated results when using similarity driven sampling. For the retrieval queries, we use EmApprox in a similar fashion to {\em distributed information retrieval}~(DIR), where a query is only processed against subcollections that are expected to be most relevant to the query~\cite{Sapprox}. Finally, for the recommendation queries, we use the user-centric collaborative filtering~(CF) algorithm~\cite{ricci2015recommender} to predict a target user's ratings using the average of other users' ratings weighted by similarities between their product reviews.

\myparagraph{Evaluation.} We generate a large number of queries for each query type, and execute them on three different data sets. We adopt equal probability cluster sampling~\cite{lohr2009sampling} over subcollections as the baseline for our evaluation. We show that EmApprox can achieve significant improvements on different domain-specific metrics
compared to the baseline with very little extra overhead during query processing. For example, to match the error bounds in aggregation queries achieved by EmApprox, the baseline would have to process $\sim$4x the amount of data. We also show that EmApprox can achieve significant speedups if users can tolerate modest amounts of imprecision. For example, when sampling at 10\%, EmApprox speeds up a set of queries counting phrase occurrences by almost 10x while achieving estimated relative errors of less than 22\% for 90\% of the queries.

EmApprox is extremely efficient for processing queries that estimate results such as aggregation and recommendation queries. In contrast, like all sampling-based approaches, EmApprox is less effective for speeding up queries similar to information retrieval queries. This is because these queries are seeking specific data items in the data set, and it is impossible to estimate missed data items based on the sample.

\myparagraph{Contributions.}~In summary, our contributions include: (i) to our knowledge, our work is the first to leverage an NLP model to build a general-purpose index to guide the approximate execution of text analytic queries; (ii) we show that the training objective of PV-DBOW is directly correlated with minimizing the variance of counting queries; (iii) we propose similarity driven sampling that can significantly increase accuracy compared to random sampling for three distinct types of approximate queries in three different application domains; (iv) we show that hashing real-valued vectors into light-weight LSH bit vectors significantly improves storage and computation efficiency without compromising precision.

%% file: related.tex
\section{Background and Related Work} 
Our work is inspired by recent work~\cite{LearnedDBIndex} that proposes using machine learned models to build database index that probabilistically map queried keys to the positions of desired records. A queried key and the matching record is analogous to an approximate query and relevant data in our work. However, our index targets approximating text analytic queries which are very different workloads than typical database queries.

\subsection{Approximate Query Processing}
Traditional AQP systems have mainly targeted aggregation queries over relational data sets. {\em AQP++}~\cite{peng2018aqp++} is a recent database system that uses sampling-based AQP and precomputed aggregates to achieve interactive response time for aggregation queries. \textit{BlinkDB}~\cite{BlinkDB} is an AQP system that selects offline-generated samples to answer queries. It uses query column sets~(QCSs) for representing the sets of columns appearing in past workloads, and stratified samples are created for each QCS. It assumes QCSs are stable over time, which does not perform well for queries outside the QCS coverage in the offline samples. \textit{ApproxHadoop}~\cite{ApproxHadoop} and \textit{ApproxSpark}~\cite{ApproxSpark} are online cluster sampling based frameworks that supports approximating aggregation with error bounds. However, their result estimation is prone to large error bounds over skewed data. \textit{Sapprox}~\cite{Sapprox} has an offline and an online component. It collects the occurrences of sub-data sets offline, and the information to facilitate online cluster sampling. \textit{Sapprox} targets certain counting queries in relational data sets, whereas EmApprox targets text analytics with a more general-purpose indexing scheme. \futurenote{\guangyannote{review later}}

\reconsider{ApproxSpark~\cite{ApproxSpark} generalizes ApproxHadoop by adapting online cluster sampling from the MapReduce computation model to multi-step RDD transformations. It also implements a distributed version of stratified sampling~\cite{ASRS} to reduce sampling errors for rare keys, but its performance gain is limited because of a larger overhead from online stratified sampling.}

\reconsider{
It was also pointed out that if user needs answers with higher accuracy but an offline sample is not large enough, then a new offline sample has to be generated which requires a full-scan of the data set similar to the cost of a precise computation. Sapprox is more flexible in that its SegMap allows online sample generation under a user-specified sampling rate but with a reduced variance
}

\subsection{Cluster Sampling}
Simple random sampling, stratified sampling and cluster sampling are three common methods of sampling, where the most computationally efficient sampling method is cluster sampling since it avoids full scan of the data set. As large data sets are usually partitioned, a cluster often would correspond to a partition~\cite{ApproxHadoop, Sapprox}.

Suppose we want to estimate the frequency $\tau$ of a phrase $Z$ occurring in a large data set partitioned into disjoint subcollections of documents. If we take a cluster sample from the data set using subcollections as the sampling units, we can derive the estimator $\hat{\tau}$ using cluster sampling theory~\cite{lohr2009sampling}:
\begin{align}
	\hat{\tau} = \frac{1}{n}\sum_{s \in S}\frac{\tau_{s}}{\phi_{s}} \pm \epsilon
	\label{eq:errorBound}
\end{align}
where $S$ is the chosen sample, $s$ is a subcollection in the sample, $\tau_{s}$ is the frequency of $Z$ in $s$,  $\phi_{s}$ is the sampling probability for $s$,
and $\epsilon$ is the estimated error bounds. $\epsilon$ in turn is computed as:
\begin{align}
	\epsilon = t_{n-1, 1-\alpha/2} \sqrt{\hat{V} (\hat{\tau})} = t_{n-1, 1-\alpha/2} \sqrt{\frac{\sum_{s \in S} (\frac{\tau_{s}}{\phi_{s}} - \hat{\tau})^2}{n(n-1)}}
	\label{eq:epsilon}
\end{align}
where $n$ is the sample size and $t_{n-1, 1-\alpha/2}$ is the critical value of a $t$-distribution at confidence level $\alpha$ with $n-1$ degrees of freedom. We observe that as the $\phi_{s}$'s approach $\frac{\tau_{s}}{\hat{\tau}}$, $\hat{V} (\hat{\tau})$ and hence $\epsilon$ will approach 0. The goal of {\it probability proportional to size}~(pps) sampling~\cite{lohr2009sampling} is to set each $\phi_{s}$ close to $\frac{\tau_{s}}{\hat{\tau}}$ by leveraging auxiliary information of each sampling unit, so that we can reduce the error bound for our estimator $\hat{\tau}$. If $\hat{\tau}$ is a good estimator, then $\phi_{s}$ should be close to $\frac{\tau_{s}}{\tau}$ which would make the estimator have a small variance.

\subsection{Paragraph Vectors} 
Recent advances in NLP have shown that semantically meaningful representations of words and documents can be efficiently learned by neural embedding models~\cite{word2vec,doc2vec}. Word2vec uses an unsupervised neural network to learn vector representations~(embeddings) for words~\cite{word2vec}. It seeks to produce vectors that are close in a vector space for words having similar contexts, which refers to the words surrounding a word in a pre-determined window. For example, synonyms like ``smart'' and ``intelligent,'' or closely related words such as ``copper''  and ``iron,'' are likely to be surrounded by similar words, so that Word2vec will produce spatially close vector representations for them. Similarity between two words can thus be scored based on the dot product distance between their corresponding vectors. \reconsider{$sim(w_1,w_2) = $. \thunote{Put in formula.} \guangyannote{I would rather not use the symbol sim here, since I defined sim = p(w|d) eq(\ref{eq:simdef}), which is proportional to $exp(\vec{w}\vec{d})$}}The learned vectors also exhibit additive compositionality, enabling semantic reasoning through simple arithmetic such as element-wise addition over different vectors. For example, $vec(``king") - vec(``man") \approx vec(``queen") - vec(``woman")$. 

Paragraph Vector~(PV)~\cite{doc2vec} is similar to Word2vec, which jointly learns vector representations for words and variable-length text ranging from sentences to entire documents in a method. Distributed Bag of Words PV (PV-DBOW) is a version of PV that has been shown to be effective in information retrieval due to its direct relationship to word distributions in text data sets~\cite{paragraphVectorIR}. By setting PV-DBOW's window size to be each of the document's length, the generative probability of word $w$ in a document $d$ is modeled by a softmax function:
\begin{equation}
P_{PV}(w|d) = \frac{exp(\vec{w} \cdot \vec{d})}{\sum_{w^\prime \in V} exp(\vec{w^\prime} \cdot \vec{d})}
\label{eq:PVSoftmax}
\end{equation}
where $\vec{w}$ and $\vec{d}$ are vector representations for $w$ and $d$, and $V$ is the vocabulary (i.e., the set of unique words in the data set). PV-DBOW
learns the word and document vectors using standard maximum likelihood estimation~(MLE), by maximizing the likelihood of observing the training text data set under the distribution defined by Eq~(\ref{eq:PVSoftmax}). As a result, the training process will output word and document vectors that satisfy Eq~(\ref{eq:PVSoftmax}) which formulates the theoretical foundation of our approximation index.

To reduce the expense of computing Eq~(\ref{eq:PVSoftmax}) during training, a technique called negative sampling has been proposed~\cite{word2vec} that randomly samples a subset of words in that document according to a noise distribution to approximate Eq~(\ref{eq:PVSoftmax}). The training process is equivalent to implicitly factorizing a shifted matrix of point-wise information~($\textit{PMI}$) between words and documents:~\cite{NeuralMatrix}:
\begin{equation}
\vec{w} \cdot \vec{d} = \textit{PMI}(w, d) - log(k)
\label{eq:wd}
\end{equation}
\reconsider{
\begin{equation}
	\textit{PMI}(x, y) = log \frac{p(x, y)}{p(x)p(y)} = log \frac{p(x|y)}{p(x)}
\end{equation}}
where $PMI(w, d)$ is the point-wise information between word $w$ and document $d$, and $k$ is a constant representing the number of negative samples for each positive instance in the training process. {\em PMI} can be estimated empirically by observing the frequencies of words in documents in the data set as $log \frac{\#(w, d)}{|d|}\cdot\frac{|D|}{\#(w,D)}$, 
where $\#(w,d)$ is the frequency of $w$ in document $d$, $|d|$ is the length of (number of words in) $d$, $D$ is the data set, $\#(w,D)$ is the total number of occurrences of $w$ in $D$, $|D|$ is the total number of words in $D$. 

Given $PMI$'s definition, Eq~(\ref{eq:wd}) reveals that the exponential of the distance between a document and word vector is proportional to the probability of document $predicting$ this word $p(w|d)$, which indicates that if a word is chosen randomly from $d$, then what is the probability that it would be $w$:
\begin{equation}
exp(\vec{w} \cdot \vec{d}) = \frac{p(w|d)}{p(w)k} \propto p(w|d)
\label{eq:wd2}
\end{equation}
We can see from Eq~(\ref{eq:wd2}) that the inner-product distance between the word vector and each document vector is proportional to the percentage of the total occurrence of $w$ contributed by each document $d$. 
\reconsider{
Negative sampling randomly samples words according to a predefined noise distribution and uses these words to approximate eq~(\ref{eq:PVSoftmax}). The global training objective of PV-DBOW using negative sampling is:
\begin{equation}
l = \sum_{(w,d) \in D}\#(w,d)~(log~\sigma(\vec{w} \cdot \vec{d})~
+ k~\cdot E_{w_N \sim P_{n}}[log~\sigma(-\vec{w_{N}} \cdot \vec{d})])
\label{eq:PVTrainingObj}
\end{equation}
where $\#(w,d)$ denotes the frequency of an observed word-document pair, $D$ is the corpus, $k$ is the number of negative samples, $\sigma$ is the sigmoid function and $E_{w_N \sim P_{v}}[log~\sigma(-\vec{w_{N}} \cdot \vec{d})]$ is the expected value of $log~\sigma(-\vec{w_{N}} \cdot \vec{d})$ given a noise distribution $P_{n}$ for $w_N$~\cite{doc2vec}.
}
\subsection{Locality-Sensitive Hashing} 
\reconsider{
As we know, index structures can make data access/search more efficient. For example in a database, B-tree based index can speed up range queries, Bloom filters are used for quickly checking record existence and hash maps makes single-key lookups $O(1)$ operations~\cite{LearnedDBIndex}. It is also pointed out in~\cite{LearnedDBIndex} that traditional index structures can be viewed as models, and using ML model to learn the distribution of keys can result in a general-purpose index which can complement existing index structures. 
}
Hashing methods have been studied extensively for searching for similar data samples in a high-dimensional data set to solve the approximate nearest neighbor problem. LSH is among the most popular choices for indexing a data set using hashing~\cite{ANNLsh}. The basic idea behind LSH is to transform each item in a high dimensional space into a bit vector with $b$ bits, using $b$ binary-valued hash functions $h_0$, ..., $h_b$. In order for the bit vector to preserve the original vectors' similarity, each hash function $h$ must satisfy the property:
\begin{equation*}
	Pr[h(\vec{x}) = h(\vec{y})] \propto sim(\vec{x}, \vec{y})
\end{equation*}
where $\vec{x}$ and $\vec{y}$ are two vectors in the data set; $sim$ is a similarity measure, such as Jaccard, Euclidean or cosine. $Pr[h(\vec{x}) = h(\vec{y})]$ is computed as one minus the ratio of the Hamming distance between two bit vectors over the total number of bits in them. Similarity between two items is preserved in the mapping, that is, if two items' LSH bit vectors are close in Hamming distance then the probability that they are close to each other in the original metric space is also high. This property allows items' LSH bit vectors to efficiently index a data set for similarity search~\cite{ANNLsh}. 

\reconsider{For example, we can use LSH to directly to rank items against a query item, exploiting the property that two bit vectors' Hamming distance can be computed efficiently. The bit vector based ranking can quickly filter out less relevant items to the query.}

%% file: technical.tex
\section{Similarity-driven Sampling}
\label{sec:methodology}

In this section, we discuss cluster sampling with probabilities driven by similarities of a query to subcollections. Cluster sampling has been adopted for approximating aggregation queries~(section 2.2) that seek to compute a sum/mean over the data set, such as counting the occurrence of a phrase, number of documents in a given topic. We show that the similarities can be computed online using the offline trained PV-DBOW vectors. Finally we describe our approximation index and using LSH to compress the real-valued vectors.

\myparagraph{Query vector representation.}~We assume a query $q$ is represented as a piece of text of $l$ words $\{w_i\}$. Under the bag-of-words assumption, the probability of $q$ in a document $d$ is the joint probability of its words $w_i$:
\begin{equation}
p(q|d) = \prod_{i \in l} p(w_i|d)
\label{eq:phraseProb}
\end{equation}
If we define $\vec{q}$ as element-wise arithmetic sum of its individual words' vectors: $\vec{q}$ = $\sum_{i=1}^{l}\vec{w_i}$, then by combining Eq~(\ref{eq:wd2}) and Eq~(\ref{eq:phraseProb}) we can derive that $p(q|d)$ is actually proportional to the exponential of $\vec{q} \cdot \vec{d}$, where Eq~(\ref{eq:wd2}) is derived from PV-DBOW's training objective:
\begin{equation}
\begin{split}
p(q|d) = \prod_{i \in l}p(w_i|d)  \propto exp(\vec{q} \cdot \vec{d})
\end{split}
\label{eq:wordsSumVec}
\end{equation}
which is the exactly the same form as Eq~(\ref{eq:wd2}) where a query only comprises a single word. Therefore by computing $\vec{q}$ this way, we can conveniently derive the probability of document predicting this query, under the assumption that the words are independent. 

\subsection{Sampling Probability Estimation}
We probabilistically define a document's $similarity$ to a query as $p(q|d)$ the probability of $d$ predicting $q$. $exp(\vec{q} \cdot \vec{q})$ in Eq~(\ref{eq:wordsSumVec}) can be used to compute $p(q|d)$ at sampling time using $\vec{q}$ and $\vec{d}$, both derived from the offline trained PV-DBOW model. Suppose we use documents as clusters for \textit {pps} sampling to estimate the quantity of $q$ throughout the data set, then we can use $exp(\vec{q} \cdot \vec{d})$ as each document's auxiliary information for setting sampling probabilities proportional to its similarity to $q$: 
\begin{equation}
\phi_{d}(q) = \frac{p(q|d)}{\sum_{d^\prime \in D} p(q|d ^\prime)} = \frac{exp(\vec{q} \cdot \vec{d})}{\sum_{d^\prime \in D} exp(\vec{q} \cdot \vec{d^\prime})} 
\label{eq:docProbWords}
\end{equation}

A large data set is often partitioned in many subcollections, and cluster sampling with those as clusters is a more efficient sampling design~\cite{ApproxHadoop,Sapprox}. Similar to sampling documents with similarity-driven probabilities, we propose to use a subcollection's vector representation to compute its similarity to the query and set sampling probabilities proportionally. Intuitively, we propose to define a subcollection' vector representation using the element-wise arithmetic mean of the vectors of the documents in it: $\vec{s} = \frac{1}{n}\sum_{d \in s} \vec{d}$, where $n$ is the number of documents in $s$, and $d$ is a document in $s$. 

We now analyze why arithmetic mean of document vectors is reasonable as subcollection's vector representation. Given $\vec{s}$ and Eq~($\ref{eq:wordsSumVec}$), we can derive the exponential of the dot product between a subcollection and query as the geometric mean of each $p(q|d)$ in $s$:
\begin{equation}
	exp(\vec{q} \cdot \vec{s}) = \sqrt[n]{\prod_{d \in s} exp(\vec{q} \cdot \vec{d})} \propto \sqrt[n]{\prod_{d \in s}  p(q|d)}\label{eq:blockSim0}
\end{equation}
Let $p(q|s)$ denote $\sqrt[n]{\prod_{d \in s}  p(q|d)}$, then we can rewrite Eq~(\ref{eq:blockSim0}) in a similar form as Eq ~(\ref{eq:wordsSumVec}):
\begin{equation}
	 p(q|s) \propto exp(\vec{q} \cdot \vec{s})
	\label{eq:blockSim}
\end{equation}
where we use $p(q|s)$ to express the {\em similarity} of a subcollection to the query. Similar to computing the similarity of $q$ to a document, we can use Eq~(\ref{eq:blockSim})'s left hand side to compute the similarity of $q$ to a subcollection. Following the same idea as using documents as sampling units, we define a probability distribution $\phi_{s}(q)$ for each subcollection $s$ with respect to $q$ in the same form as Eq~(\ref{eq:docProbWords}):
\begin{equation}
	\phi_{s}(q) = \frac{p(q|s)}{\sum_{s^\prime \in D} p(q|s^\prime)} = \frac{exp(\vec{q} \cdot \vec{s})}{\sum_{s^\prime \in D} exp(\vec{q} \cdot \vec{s^\prime})} 
	\label{eq:blockProbWords}
\end{equation}
where $exp(\vec{q} \cdot \vec{s})$ can be computed at sampling time.

A cluster sample over the subcollections with probabilities set according to Eq~(\ref{eq:blockProbWords}) will greatly reduce the uncertainty in estimating the occurrence of $\{w_i\}$. Variants of this query include estimating number of documents that contain $\{w_i\}$ or number of documents similar to $\{w_i\}$ in semantics. Both distributions defined in Eq~(\ref{eq:docProbWords}) and (\ref{eq:blockProbWords}) essentially normalize the probability of a phrase appearing in a specific document or subcollection, against every document or subcollection in the data set. Interestingly, Eq~(\ref{eq:docProbWords}) and (\ref{eq:blockProbWords}) have the same form as a softmax classifier over query words $\{w_i\}$ which predicts its probabilities conditioned on a document or subcollection.

\reconsider{
	It mentions that the lengths of vectors are related to the word significance.
	\subsubsection{Two-stage sampling}
	Based on the similarity between query and subcollection vectors, relevant subcollections can be first sampled depending on their similarity to the query. After choosing the subcollections, the individual documents in the chosen blocks can be used as secondary sampling units. The document vectors can be used to sample documents proportional to their similarities to the query. 
	
	This sampling design corresponds to two-stage unequal probability sampling where each stage takes sample with probabilities proportional to similarities.
}

\subsection{Approximation Index}
The approximation index includes vectors for every word, document and subcollection. The index can occupy significant storage space for a large data set, for which we propose to map the real-valued vectors to LSH bit vectors to reduce the required storage. And, the cost of computing the similarities between bit vectors is also much more efficient than dot product between real-valued vectors.

\myparagraph{LSH.}~LSH bit vectors can preserve different distance metrics~(e.g., cosine) between their real-valued vector counterparts. Computing the Hamming distance of the LSH bit vectors using {\tt XOR} is also much more efficient than dot product. We slightly modify the gradient descent-based training process of PV-DBOW: at each update step, we normalize the vectors to be unit length so that the dot product of the trained vectors is equivalent to the cosine similarities between two vectors. The value of the hash function for preserving cosine distance depends on the dot product between a random plane $\vec{r}$ and an item vector $\vec{x}$, where $h_r(\vec{x})$ evaluates 1 if $\vec{r} \cdot \vec{x} \geq 0$, and 0 otherwise, 
where $\vec{r}$ usually has a standard multi-dimensional Gaussian distribution $\mathcal{N}(\mathbf{0}, I)$, and a new $\vec{r}$ is generated each time the hash function is applied~\cite{ANNLsh}. In order to generate the LSH signature for a real value vector, we first choose a dimension $l$ for the bit vector, then apply hash function $h_r(\vec{x})$ $l$ times to generate each bit, each choosing a random $\vec{r}$. Specifically, we can approximate $exp(\vec{w} \cdot \vec{d})$ using $exp(cos(\frac{m}{l}\pi))$, where $m$ is the Hamming distance between $\vec{w}$ and $\vec{d}$'s corresponding LSH bit vectors.

%% file: tasks.tex
\reconsider{
	It mentions that the lengths of vectors are related to the word significance.}

\section{Beyond Aggregation Query} \label{sec:tasks}
\subsection{Query characterization}
In the previous section, we introduced approximating aggregation queries such as estimating occurrence of a phrase. In this section, we describe approximating DIR and recommendation queries using EmApprox. We note that the goal of IR is to identify ``similar'' documents to a query, and that some recommendation technique in data mining depend on identify ``similar'' users to a target user. We characterize our targeted queries as: 1) the query can be represented by words; 2)~relevancy of the query to a subset of data can be reduced to the generative probability of the query's relevant data given this subset - the definition of similarity; 3)~efficiency of the approximate computation can be improved by processing the most similar data in the data set.

\subsection{Distributed Information Retrieval} \label{sec:dir}
\reconsider{In IR, a score is usually computed for each document indicating how relevant it is to the user's query~(usually a sequence of words), to retrieve the relevant documents. The score can be computed based on metrics such as \textit{tf-idf} of the query words or the generative probability of the query given a document using a language model, known as query likelihood~\cite{wei2006lda}. IR can benefit from representations of words and documents in vectors that encode semantic information~\cite{paragraphVectorIR}, where scoring of the documents can be computed based on the similarity between a query and documents. 
A retrieval model proposed in~\cite{wei2006lda} uses latent Dirichlet allocation~(LDA) for query likelihood. LDA is a generative language model that abstracts a document as mixture of topics, represented by a stochastic vector $\theta_{d} \in \Delta ^{K}$, where $K$ is the number of topics. Inspired by~\cite{wei2006lda}, a Paragraph Vector-based retrieval model was proposed in~\cite{paragraphVectorIR} that has $P_{PV} (w|d)$ defined in eq~(\ref{eq:PVSoftmax}) in its retrieval model.

\begin{equation}
P(w|d) = (1-\lambda) P_{QL}(w|d) + \lambda P_{PV} (w|d)
\end{equation}
where $P(w|d)$ is the probability of $w$ given $d$, $P_{QL}(w|d)$ is defined as $P_{PV}(w|d)$ is defined in eq~(\ref{eq:PVSoftmax}), $\lambda$ is a free parameter between 0 and 1.
}

Information retrieval from disjoint subcollections of documents is known as distributed information retrieval~(DIR), where many irrelevant subcollections are ignored for improved retrieval efficiency~\cite{kulkarni2015selective}. EmApprox can facilitate subcollection selection under the vector space retrieval paradigm for DIR~\cite{croft2015search, kulkarni2015selective}. We target Boolean and ranked retrieval models for DIR in the following discussion. 
\reconsider{However, our goal is not to compete with existing DIR solutions, but to demonstrate that our proposed index can complement previous AQP systems and extend approximation tasks to DIR with efficiently.}

\myparagraph{Boolean retrieval.}~Under the Boolean model, a query is a Boolean expression of words connected by Boolean operators~(e.g., $w_0 \lor (w_1 \land w_2)$), where a term $w_i$ only evaluates to true when contained in a document~\cite{croft2015search}. The retrieval result is a set of documents that satisfy the Boolean expression.

When approximating a Boolean query $q_{b}$, we first compute its similarity to each subcollection - $p(q_{b}|s)$ as we have previously defined. We can compute $p(q_{b}|s)$ using the same sequence as how the Boolean query is evaluated, i.e. $\land$ takes precedence over $\lor$. We first compute each individual term~($w_i$)'s $p(w_i|s)$ to a subcollection $s$ using Eq~(\ref{eq:blockSim}), in order to compute the query's overall similarity. Since $w_i \land w_j$ implies that $w_i$ and $w_j$ both have to exist for it to be true, whereas satisfying $w_i \lor w_j$ requires either $w_i$ or $w_j$ to exist, the generative probability of $w_i \land w_j$ is equivalent of $p(w_i|s) \cdot p(w_j|s)$; by the same token, the generative probability of $w_i \lor w_j$ is therefore $p(w_i|s) + p(w_j|s)$. For example, suppose we have a Boolean query $q_{b} = w_0 \lor (w_1 \land w_2)$, then $p(q_{b}|s)$ can be computed as $p(w_0|s) +  (p(w_1|s) \cdot p(w_2|s))$, where each $p(w_i|s)$ is computed using Eq~(\ref{eq:blockSim}). Finally, we use the overall query $similarity$ to the subcollections $p(q_b|s)$ to compute sampling probability of each subcollection to sample a subset of subcollections. Then, only documents in the chosen subcollections are evaluated against the Boolean query to retrieve matching documents.

\myparagraph{Ranked retrieval.}~Instead of returning a set of documents that precisely match a Boolean expression, ranked retrieval returns a list of documents ranked by their relevancy to the query. The query is usually a phrase comprising of a sequence of words. In terms of ranking, each document is assigned a score using a function with potentially many factors considered, such as the \textit{tf-idf} of the query terms or cosine similarity between the vector representations of the query and a document obtained under the same language model. Similar to Boolean retrieval, our proposed framework first samples a subset of subcollections with probabilities proportional to the query's $similarity$ to each subcollection, which is more straightforward to compute than Boolean model - we just use Eq~(\ref{eq:blockProbWords}) to compute its $similarity$ to a subcollection. We then apply a user-specified scoring function to documents from the chosen subcollections, such as BM25~\cite{bm25} or a language model-based function~\cite{paragraphVectorIR}.

\subsection{Recommendation} \label{sec:rec}
\reconsider{Recommender systems generally predict a user's response to unrated items based on factors such as user past behavior and/or properties of the items, and there is a large body of previous work on recommendation techniques~\cite{ricci2015recommender}.} One of the most successful recommendation techniques in data mining is called neighborhood-based collaborative filtering~(CF), which have been deployed in many commercial websites~\cite{ricci2015recommender, ning2015comprehensive}. User-centric neighborhood CF is that given a target user $u$, it first tries to identify a subset of more similar users as \textit{neighbors}, then uses the similarity score between $u$ and each neighbor to predict a rating for item $i$. It computes the prediction by taking an average of all the ratings for $i$ by $u$'s neighbors weighted by their similarity scores. Symmetrically, neighborhood CF can be item-centric which predicts a rating for an item $i$ based on similarities between $i$ and other items purchased by $u$. 

Without loss of generality, we will focus on user-centric CF. Review text embeds rich information which has been used to model users' behaviors and items' properties in past work as an alternative to numerical rating only data~\cite{McAuley:2013}. A user's vector representation can be learned under PV-DBOW model by defining a document as all the reviews a user has written~\cite{McAuley:2013}. Consequently, a user's vector $\vec{u}$ would encode $u$'s preference. When the number of total users in the data set is large, the process of identifying neighbors from them to a target user $u$ would be expensive. Our proposed index can make it more amenable for large data sets by selecting only more similar subcollections of users. Suppose the review data set is sorted by users, then similarity of a user $u$ to a collection of reviews can be computed using Eq~(\ref{eq:blockSim}) to sample most similar collections of users. Then predicted ratings for the new user can be computed using any model/metric with the neighbors in the chosen subcollections. 

As a concrete example, rating of an item $i$ by user $u$'s predicted value can directly leverage $u$'s review text vector's similarity to a user $v$ who has also rated item $i$, computed as $\sum_{v \in U^\prime} sim(u,v)r(v,i)$, where $U^\prime$ is the set of all users who have rated item $i$ in the chosen subcollections, $r(v,i)$ is user $v$'s rating for item $i$, $sim(u,v)$ can be characterized by their similarity in the review texts $u$ and $v$ have written, defined as $\sum_{v \in U^\prime}\frac{exp(\vec{u} \cdot \vec{v})}{\sum_{v^\prime \in U^\prime}exp(\vec{u} \cdot \vec{v^\prime})}$.


\reconsider{
$\vec{u}$ and $\vec{v}$ have two parts, the first part is the past ratings to all products and the second part is the review text vector. For unrated items, the rating will be zero. \yongfengnote{why would we integrate the rating vector here, does it still satisfy the good mathematical properties derived above?}  
}
\reconsider{
\subsubsection{Content filtering.} 
\yongfengnote{I guess you mean item-based collaborative filtering in this subsection? because content-based filtering does not mean what you described here.}
Neighborhood methods translate either users or items into the same vector space and compare their similarities. Content-based methods relates user to items and compare their similarities in the same vector space. 

Similarly, the rich information in review text can still be leveraged in a similar way like the neighborhood methods. In this case, we assume the data set containing reviews is sorted by item Ids. The predicted rating of user for a new item is thus the average rating by the user, weighted by the similarities of this new item to each of the item in this user's purchase history. 
}
\futurenote{
\subsection{Discussion}
We hypothesize that using documents as sampling units is better for information retrieval-oriented tasks where subcollection-based cluster sampling may be too coarse. It is because IR is very sensitive to the allocation of documents in the data sets. If semantically documents are not clustered together, dropping subcollections may result in missing highly similar documents to the query. 

However, using subcollection as sampling units is preferred since computing closed-form estimator to the query is less prone to losing particular documents. Sampling subcollections is more efficient because it eliminates the need of I/O for the dropped subcollections. 
}

\reconsider{We define two categories for our target approximation tasks: one is estimating a quantity using aggregation, the other is retrieving relevant information that match a query. Aggregation outputs a confidence interval for an estimator and will have a smaller error bound by using {\it SubSim} as auxiliary information for {\it pps} sampling. In the retrieving data case such as DIR, {\it SubSim} directly indicates the a subset of data's similarity to a query.} 

\subsection{Discussion on Document Allocation}\label{sec:clustering}
The document allocation policy can affect DIR's performance, where the storage of documents needs to be skewed for DIR to be effective~\cite{ kulkarni2015selective}. This is because the query expects to retrieve as many relevant documents to the query as possible from only a subset of the data set. The document allocation policy can also affect recommendation's performance, since identifying similar users is similar to retrieving relevant documents. On the other hand, the accuracy of aggregation estimators is not dependent on document allocation, since the local sum of each chosen subcollection is multiplied by the inverse of its own sampling probability as a scaling factor to compute an overall estimator, i.e., subcollections with a large local sum will just have a small scaling factor.

We propose to allocate documents based on their vectors' pair-wise cosine distance through clustering the documents in the original data set using \textit{spherical K-means}~\cite{zhong2005efficient}, that uses the cosine as distance metric. The clustering process takes as input the collection of document vectors, and produces an allocation where semantically similar documents are clustered.  

When documents $d_1$, $d_2$, ..., $d_n$ are semantically similar, it indicates that the probabilities $p(w|d_1)$, $p(w|d_2)$, ..., $p(w|d_n)$ are also similar for a query word $w$. The result of clustering is a more skewed distribution of the documents, therefore documents that have the same probability of predicting a query word tend to be allocated together. As Eq~(\ref{eq:blockSim0}) shows that $p(w|s) = \sqrt[n]{\prod_{d^ \prime \in s} p(w|d^ \prime)}$ -- the geometric mean of each document's probability of predicting $w$ in that subcollection $s$ -- so if $p(w|d^ \prime)$'s are similar in each $s$, then their geometric mean $p(w|s)$ will approach a local maximum equivalent to the arithmetic mean of all the $p(w|d^ \prime)$, according to the AM-GM inequality~\cite{hirschhorn2007gm}. It therefore suggests that this allocation policy would produce a skewed sampling probability distribution $\phi_{s}(w)$, which is desired for a retrieval query.

%% file: discussions.tex
\section{Limitations}
EmApprox can approximate a range of text analytical queries, we nevertheless highlight a couple of limitations.

\myparagraph{Model drift.}~We assume the text data set is historical and stable. If new documents are added to the existing data set, PV-DBOW is able to infer the vectors for an unseen document using the words in the new document~\cite{doc2vec}. However, the originally trained PV-DBOW model may drift due to document updates. Therefore PV-DBOW model should be retrained to capture the true word/frequent distributions in the data set, which requires the offline index to be rebuilt.

\myparagraph{Unseen words in the query.}~Currently we assume the query does not include words outside the vocabulary of the data set, so that any word vector in the query can be directly obtained.

%% file: implementation.tex
\section{Implementation}
\label{sec:system}

\begin{figure}[!tbp]
    \begin{center}
	    \includegraphics[width=6.5cm]{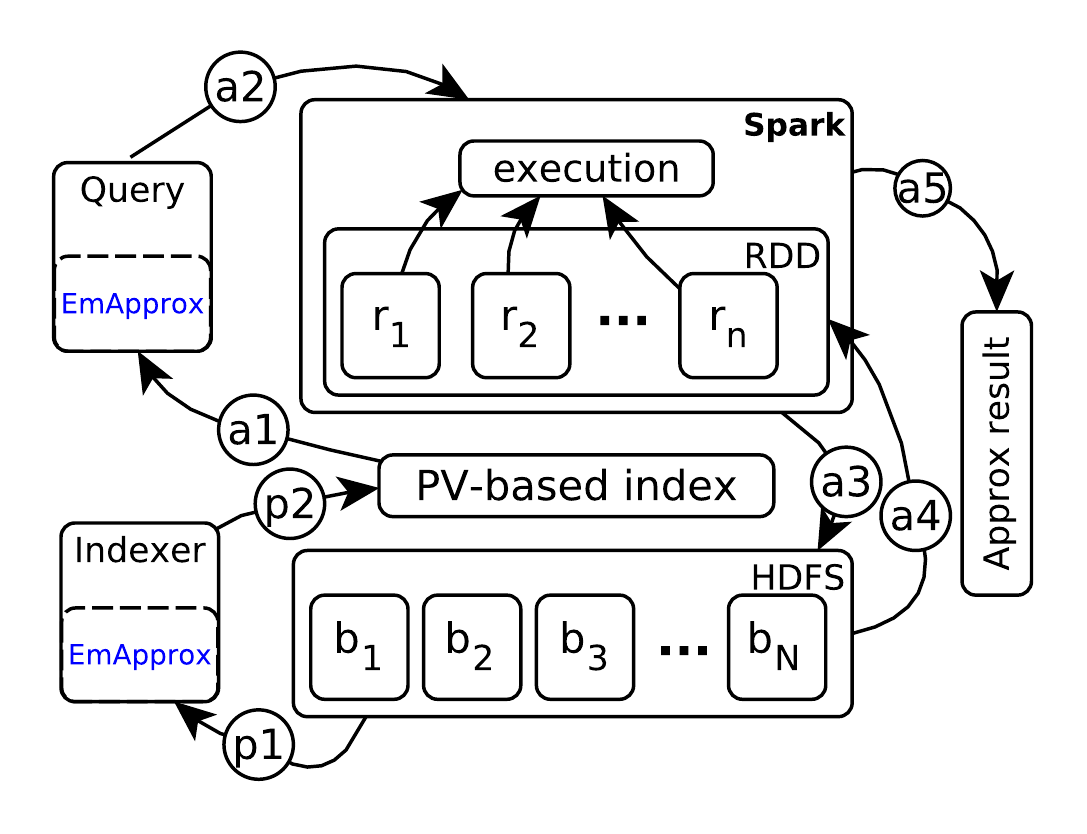}
    \end{center}
    \vspace{-0.2in}
	\caption{EmApprox architecture. $\{b_N\}$ are HDFS blocks, $\{r_n\}$ are Spark RDD partitions, where $n$ is a subset of $N$. The offline indexer and query both use the EmApprox library.}
	\label{fig:systemDesign}
\end{figure}

We have implemented a prototype of EmApprox as a Python library comprising two parts: one for building offline indexers and one for building approximate query processing applications (queries for short). Users write indexers and queries as Spark programs in Python using the PySpark~\cite{PySpark} and EmApprox libraries. Figure~\ref{fig:systemDesign} gives an overview of a system built using EmApprox, where documents are stored in blocks of an HDFS filesystem, with blocks considered subcollections of documents and used as sampling units. Note that a single indexer can be used to index many different data sets of the same type, and many different queries can be executed against each index/data set.

We leave the task of writing indexers to the user because it allows the flexibility for indexing many different types of data (e.g., different data layouts and definitions of documents). It is quite simple to write indexers given the EmApprox library. Specifically, users need to write code to parse a given data set to extract the documents (much of this code can come from standardized libraries) and identify documents in each subcollection (HDFS block). All other functionalities are implemented in the EmApprox library, and simply requires the user program to call several functions. Similarly, the main difference between an approximate query built using EmApprox and a precise query is the invocation of several EmApprox functions.

An offline indexer uses EmApprox to learn vector representations for words and documents, cluster documents (when desired) using K-means as discussed in Section~\ref{sec:clustering}, compute vectors for blocks (subcollections), compute corresponding LSH bit vectors, and prepare the index. This process is shown as steps $p1$ and $p2$ in Figure~\ref{fig:systemDesign}. (We do not show the clustering for simplicity.) We use Gensim~\cite{Gensim} as the default library for PV-DBOW model training, but we can also use alternative implementations that can run on distributed frameworks such as Tensorflow~\cite{abadi2016tensorflow} to reduce the training time. We use Gensim in our prototype because it is a widely adopted PV-DBOW implementation.

The execution of an approximate query is shown as steps $a1$ through $a5$ in Figure~\ref{fig:systemDesign}. In step $a1$, the query uses EmApprox to read the index into an in-memory hash table, compute sampling probabilities for all HDFS blocks, and choose a sample using {\it pps} sampling. Step $a2$ launches a Spark job. Steps $a3-4$ are part of the Spark job and use EmApprox and PySpark to read the sample from the data set into an RDD. Step $a5$ is the execution of the rest of the Spark job. We provide two simple reduce functions that compute the estimated sum and average, along with the confidence interval (see Eq~(\ref{eq:errorBound}) and (\ref{eq:epsilon})).

%% file: evaluation.tex
\reconsider{
\begin{table*}
	\begin{center}
		\begin{tabular}{|l|r|p{6cm}|l|r|r|}
			\hline
		    &  & &  &\bf \multicolumn{1}{c}{Training}  &\bf \multicolumn{1}{c}{Index} \\
			\bf \multicolumn{1}{c}{Dataset} & \bf \multicolumn{1}{c}{Size} & \bf \multicolumn{1}{c}{Description} & \bf \multicolumn{1}{c}{Domain}  &\bf \multicolumn{1}{c}{Time}  &\bf \multicolumn{1}{c}{Size} \\
			\hline
			Wikipedia& 62GB & $\sim$5M Wikipedia articles in XML format. \thunote{When was this snapshot taken?}  & Aggregation, DIR  & 4.3h & 125~MB\\
			\hline
			CCNews     & 65GB & $\sim$22M news articles crawled from the Web in 2016, in JSON format. & DIR   & 6.2h & 280 MB\\
			\hline
			Amazon  reviews  & 55GB & $\sim$142M user product reviews in the time period 5/1996--7/2014 in JSON format. & Recommendation   &3.8h   & 88 MB  \\
			\hline
		\end{tabular}
	\end{center}
	\caption{Evaluation data sets. Training Time is the time to train PV-DBOW. Index sizes are after conversion of learned vectors to bit vectors using LSH.}
\end{table*}
}

\begin{table*}
\fontsize{9}{9}\selectfont

	\begin{center}
		\begin{tabular}{|c|c|p{7cm}|p{4cm}|}
			\hline
			\multicolumn{1}{|c|}{\textbf{Query}} & \multicolumn{1}{|c|}{\textbf{Domain}} & \multicolumn{1}{|c|}{\textbf{Description}} & \multicolumn{1}{|c|}{\textbf{Domain-specifc metrics}}                   \\
			
			\hline
			phrase occurrence & aggregation & estimates frequency for target phrase. & error bound             \\
			\hline
			Boolean retrieval     &DIR & retrieves a (sub)set of documents that precisely match a Boolean query. & recall\\
			\hline
			ranked retrieval  & DIR & retrieves top-k documents ranked by a given scoring function over a set of query terms. &precision          \\
			\hline
			user-centric CF & recommendation & predicts ratings on unbought products and outputs top-k recommendations for a user.& MSE, precision \\
			\hline
		\end{tabular}
	\end{center}
	\caption{Approximation queries and metrics summary}
	
	\label{tbl:tasks}
\end{table*}

\begin{table*} 
\fontsize{9}{9}\selectfont
	\begin{center}
		\begin{tabular}{|c|p{4cm}|c|c|p{3cm}|c|c|}
			\hline
		    	\multicolumn{1}{|c|}{\textbf{Data set}} & 	\multicolumn{1}{|c|}{\textbf{Description}} & \multicolumn{1}{|c|}{\textbf{Queries}}  & \multicolumn{1}{|c|}{\textbf{Size}} &\multicolumn{1}{|c|}{\textbf{Document}}   &\multicolumn{1}{|c|}{\textbf{T-Time}}  &\multicolumn{1}{|c|}{\textbf{Idx-size}} \\
			\hline
			Wikipedia& $\sim$5 million Wikipedia articles in XML format. & aggregation & 62GB & each Wikipedia article  & 4.3h & 125MB\\
			\hline
			CCNews &$\sim$22 million news articles crawled from the Web in 2016, in JSON format. & aggregation, DIR & 65GB & each news article   & 6.2h & 280MB\\
			\hline
			Amazon  reviews  & $\sim$142 million user-product interactions from 5/1996- 7/2014 in JSON format. & recommendation &55GB & all reviews written by the same user   &3.8h   & 87.5MB  \\
			\hline
		\end{tabular}
	\end{center}
	\caption{Dataset descriptions, notion of a document in the data set~(Document) for PV-DBOW training, PV-DBOW training time~(T-Time) and index sizes after converted to LSH.}
	\label{tbl:data sets}
\end{table*}

\section{Evaluation} \label{sec:evaluation}

We evaluate EmApprox using synthetic queries of the three types discussed above. Table~\ref{tbl:tasks} summarizes the query types and the evaluation metrics. Table~\ref{tbl:data sets} summarizes the data sets that we use in our evaluation.

\subsection{Evaluation Methodology and Metrics}

\myparagraph{Data.}~We use three data sets: a snapshot of Wikipedia~\cite{Wikipedia}, a news corpus from Common Crawl~(CCNews)~\cite{CCNews}, and a set of Amazon user reviews~\cite{McAuley:2013}. Table~\ref{tbl:data sets} summarizes the data sets, their use in different queries, the training time of PV-DBOW using Gensim, and the size of the resulting indices after compression using LSH.

\myparagraph{Experimental platform.}~Experiments are run on a cluster of 6 servers interconnected with 1Gbps Ethernet. Each server is equipped with a 2.5GHz Intel Xeon CPU  with 12 cores, 256GB of RAM, and a 100GB SSD. Data sets are stored in an HDFS file system hosted on the servers' SSDs, configured with a replication factor of 2 and block size of 32MB. Applications are written in Python 3 and run on Spark 2.0.2.

\myparagraph{Index construction.}~We train PV-DBOW to produce word and document vectors with 100 dimensions. We use LSH vectors of 100 bits, which compresses the PV-DBOW learned vectors by a factor of 64. We explore the sensitivity of EmApprox to these parameters in Section~\ref{sec:sensitivity}.

\myparagraph{Baselines.}~We analyze and compare EmApprox's performance against simple random clustered sampling (SRCS)~\cite{lohr2009sampling} of HDFS blocks and precise execution. We compare execution times (speedups) and query-specific metrics. We run the precise executions as ``pure'' Spark programs on an unmodified Spark system. We run SRCS using the EmApprox prototype, replacing pps sampling with simple random sampling.

\myparagraph{Aggregation queries.}~We run aggregation queries that estimate the numbers of occurrences and the corresponding relative errors of target phrases in the Wikipedia data set. We create 200 queries by randomly selecting phrases from the data set. The lengths of the phrases follow a normal distribution with a mean of 2 words and a standard deviation of 1.

The approximate answer to each query executed with a given sampling rate includes the estimated count ($\hat{\tau}$) and an estimated confidence interval~($\hat{\tau}\pm \epsilon$). We report the {\em estimated relative error} at 95\% confidence level as the ratio of $\epsilon$ over $\hat{\tau}$. We also compare the estimated relative error with the actual relative error, computed as $\frac{|\hat{\tau} - \tau|}{\tau}$, where $\tau$ is the precise answer. 

\myparagraph{DIR queries.} We cluster documents within the Wikipedia and CCNews data sets as explained in Section~\ref{sec:clustering}, setting the number of centroids equal to the number of HDFS blocks in the file holding the data set.

We generate 200 sets of randomly chosen words, 100 from the Wikipedia data set and 100 from CCNews. Set sizes follow a normal distribution with an average size of 3 and a standard deviation of 1. We randomly insert Boolean operators ({\tt and} and {\tt or}) to form Boolean queries, and use the sets of words directly as queries in ranked retrieval. We use the BM25 ranking function~\cite{bm25} in ranked retrieval. We choose BM25 from a plethora of ranking functions, including functions that use Paragraph Vector~\cite{paragraphVectorIR}, because it is widely adopted by search platforms such as Solr~\cite{Solr} and IR libraries such as Apache Lucene~\cite{Lucene}. In Boolean retrieval, we report {\em recall}, defined as ratio of the number of documents retrieved by the approximate query processing to the number of documents retrieved by the precise execution of the query. In ranked retrieval, we report {\em precision-at-k} (P@k), defined as the percentage of the top $k$ documents retrieved by the approximate query processing that is in the top $k$ retrieved by the precise query execution. We report precision instead of recall in ranked retrieval because users typically cares most about the top $k$ ranked answers, where $k$ is typically small~\cite{croft2015search}.
\begin{figure*}[t]
	\subfloat[1\% Sampling Rate] {\includegraphics[width=0.25\linewidth]{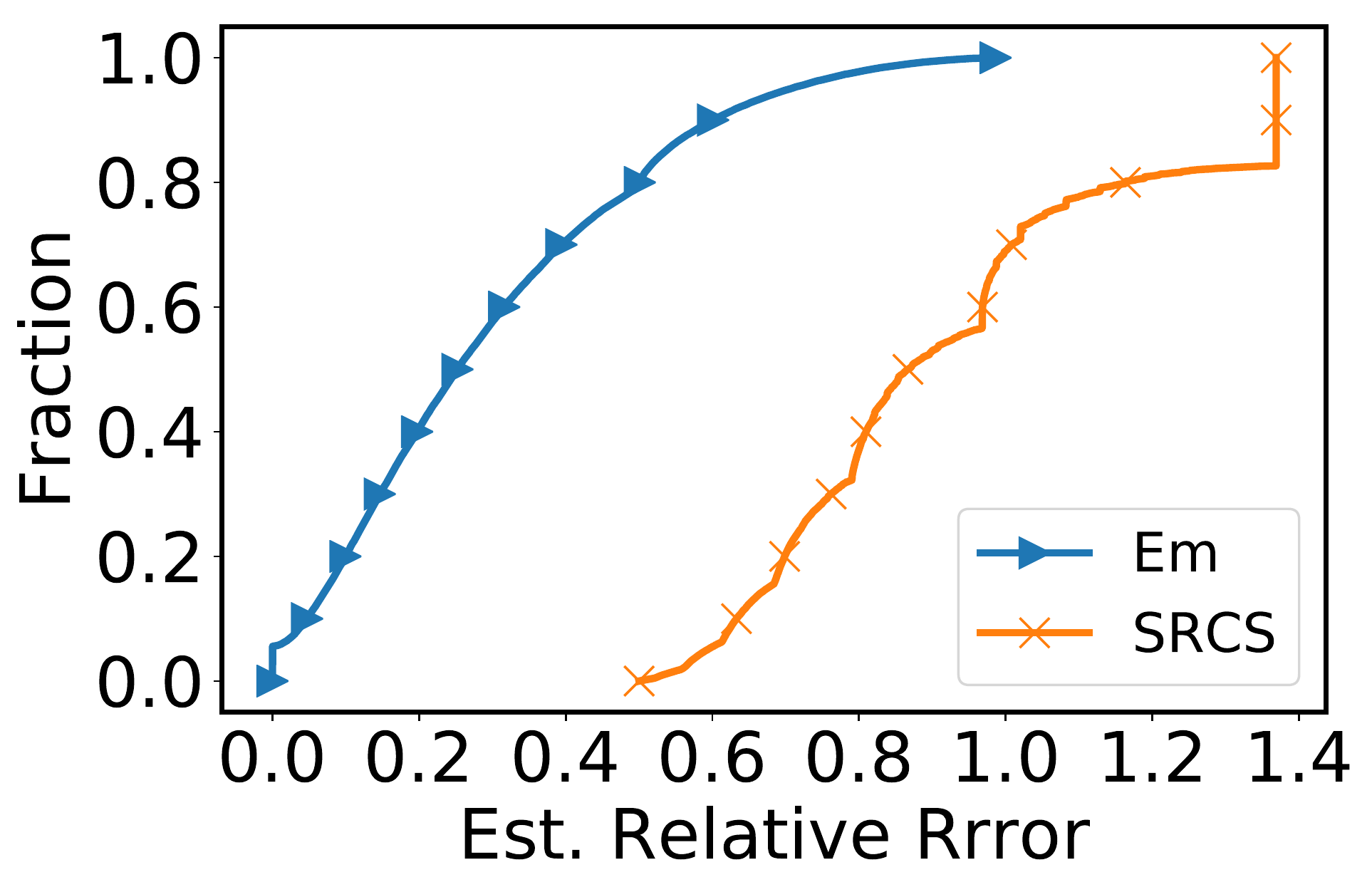}}
	\subfloat[2.5\% Sampling Rate] {\includegraphics[width=0.25\linewidth]{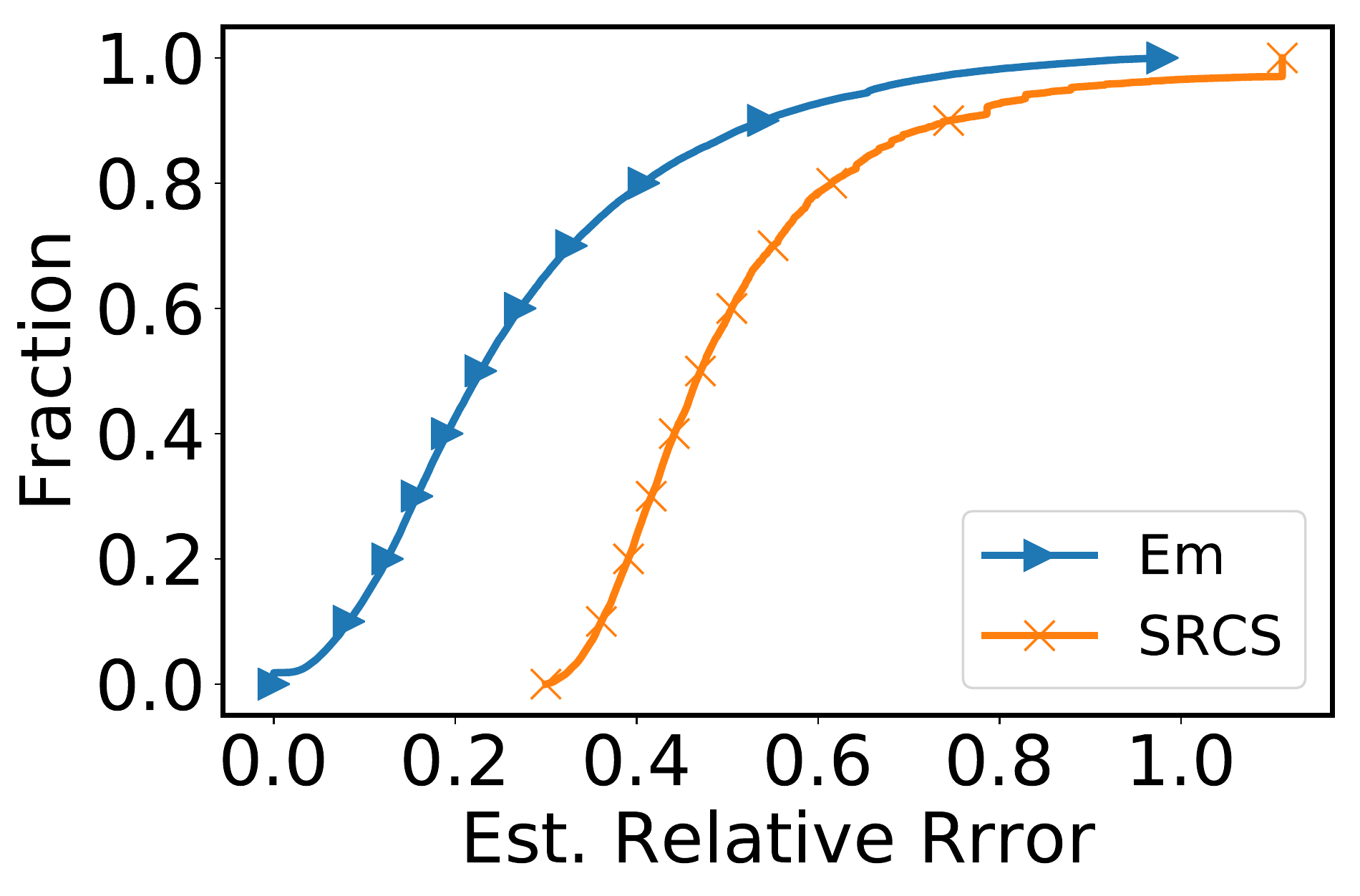}}
	\subfloat[5\% Sampling Rate] {\includegraphics[width=0.25\linewidth]{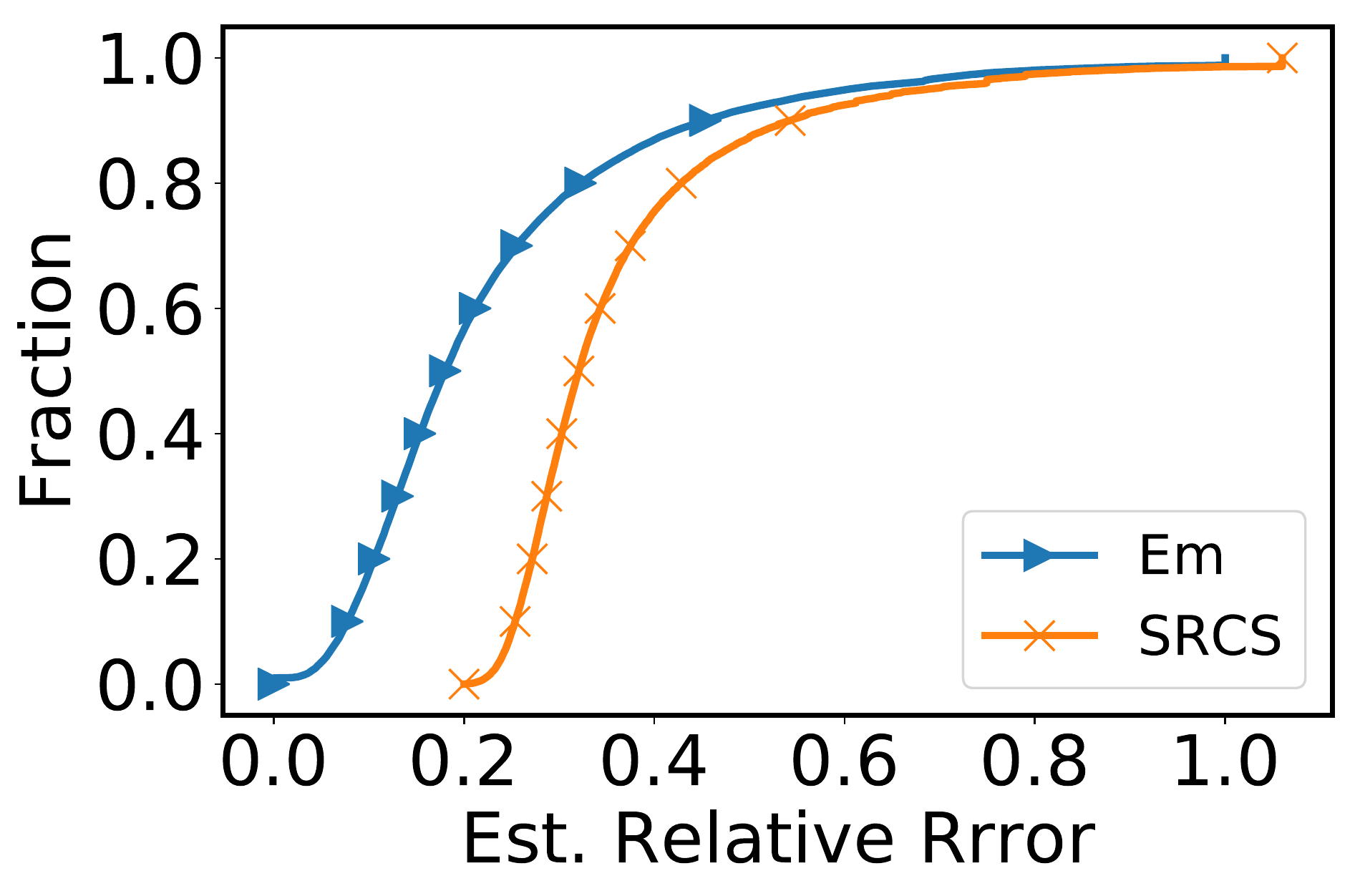}}
	\subfloat[10\% Sampling Rate]{\includegraphics[width=0.25\linewidth]{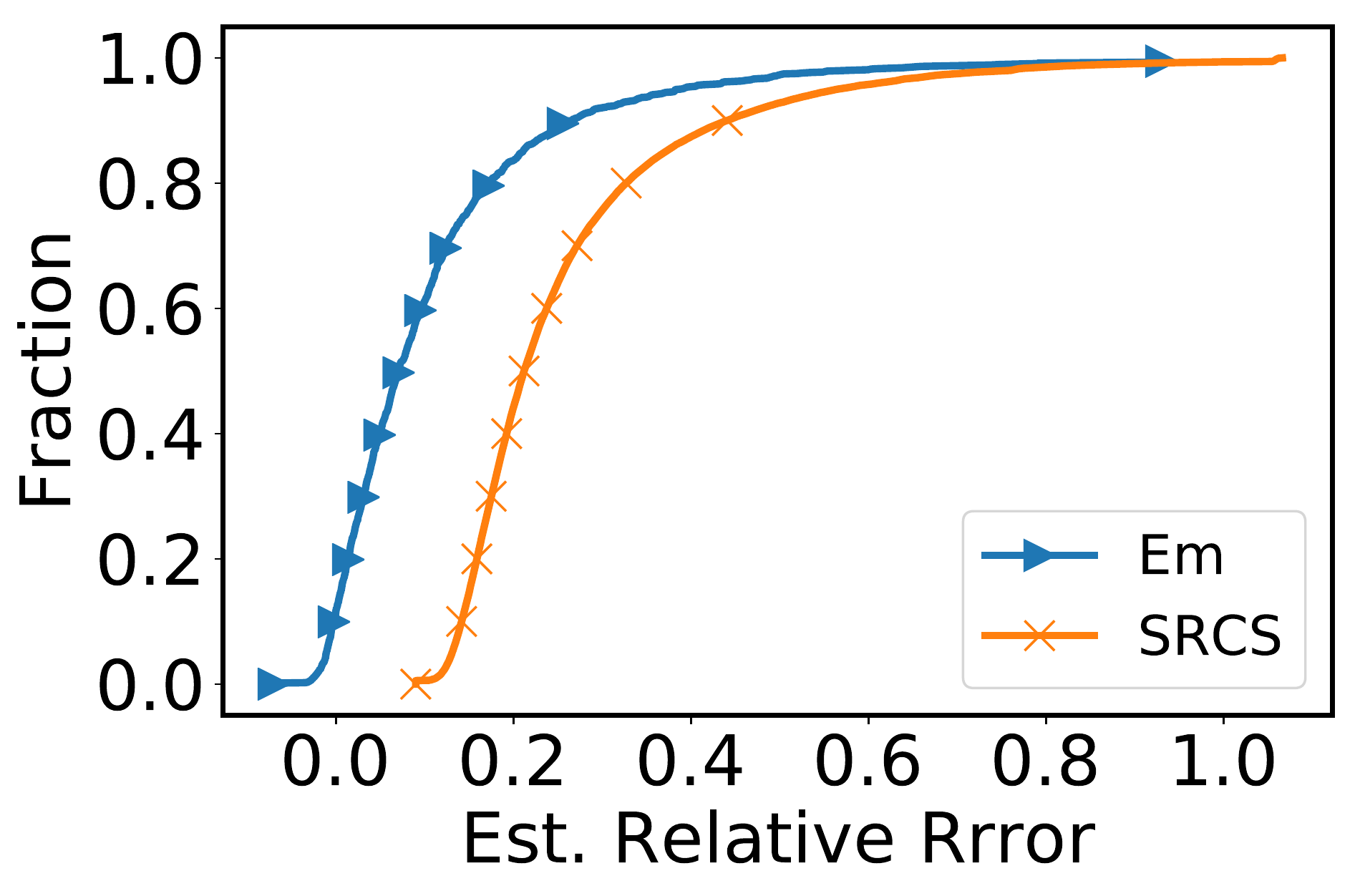}}
	\caption{CDFs of estimated relative error for phrase occurrences under different block sampling rates.}
	\label{fig:wcSamplingError1}
\end{figure*}

\begin{figure}
 \vspace{-0.2in}
	\subfloat[Avg. Speedup] 
	{\includegraphics[width=0.5\linewidth]{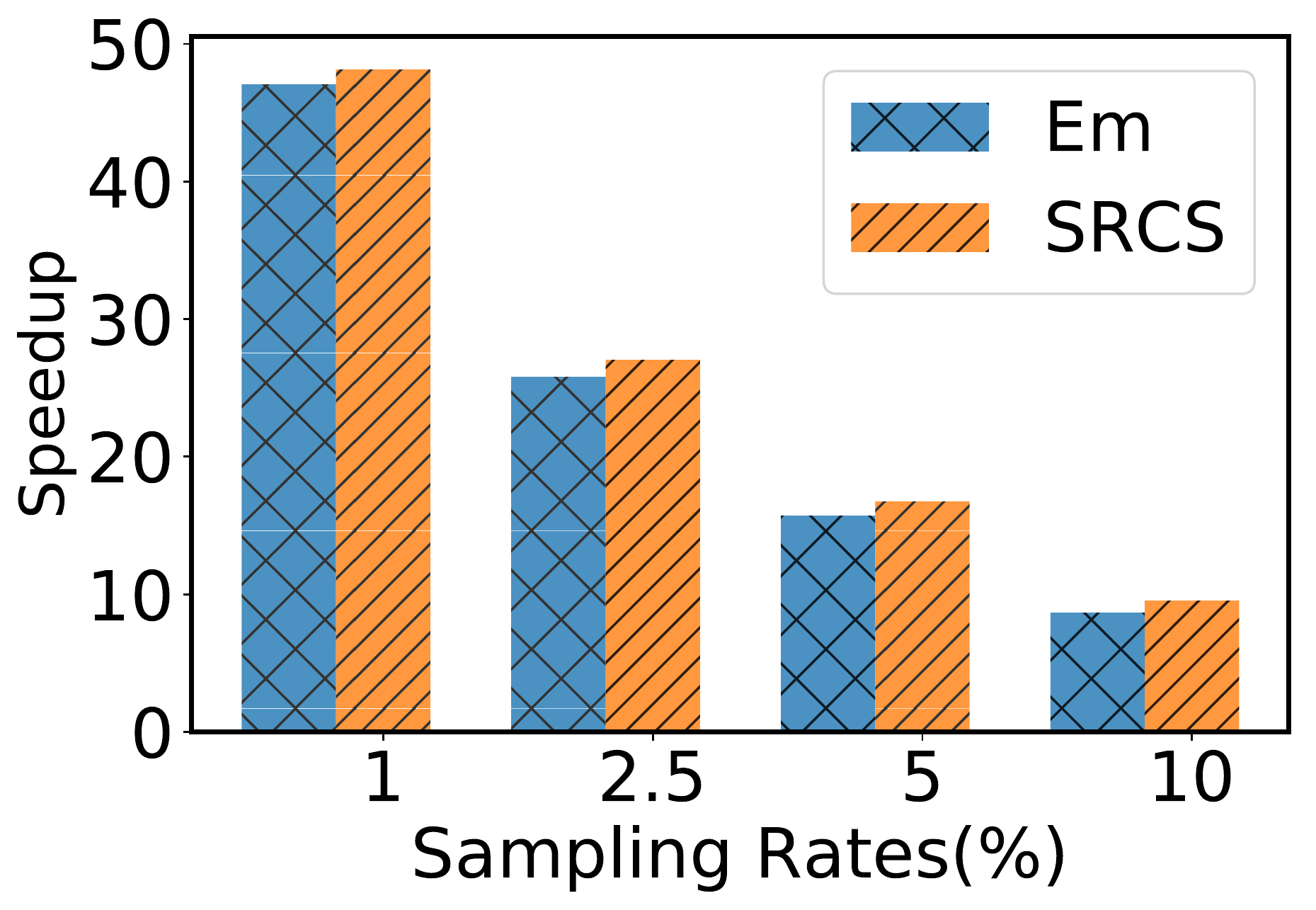}}
	\subfloat[Actual and estimated rel. errors]
	{\includegraphics[width=.5\linewidth]{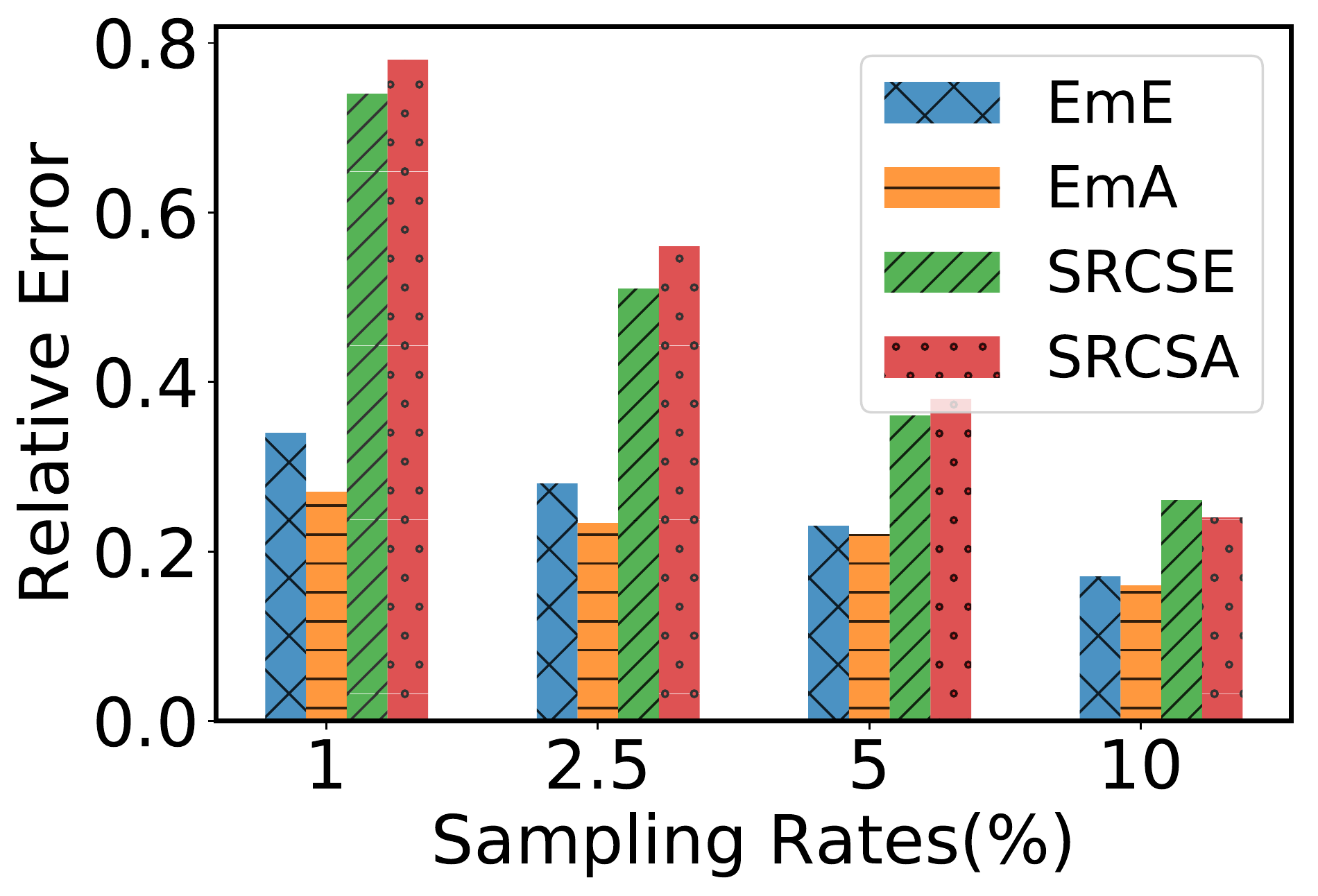}}
	
	\caption{Average speedups and relative errors for phrase occurrence query. (b) shows the comparison of estimated~(E) and actual average relative error~(A) using EM and SRCS~(e.g. EmE means estimated relative error using EmApprox, SRCSA means actual relative error using SRCS).}
	\label{fig:wordCountAvgs}
\end{figure}

\begin{figure*}[!tbp]
	\subfloat[Wikipedia, 25\% sampling]
	{\includegraphics[width=0.25\linewidth]{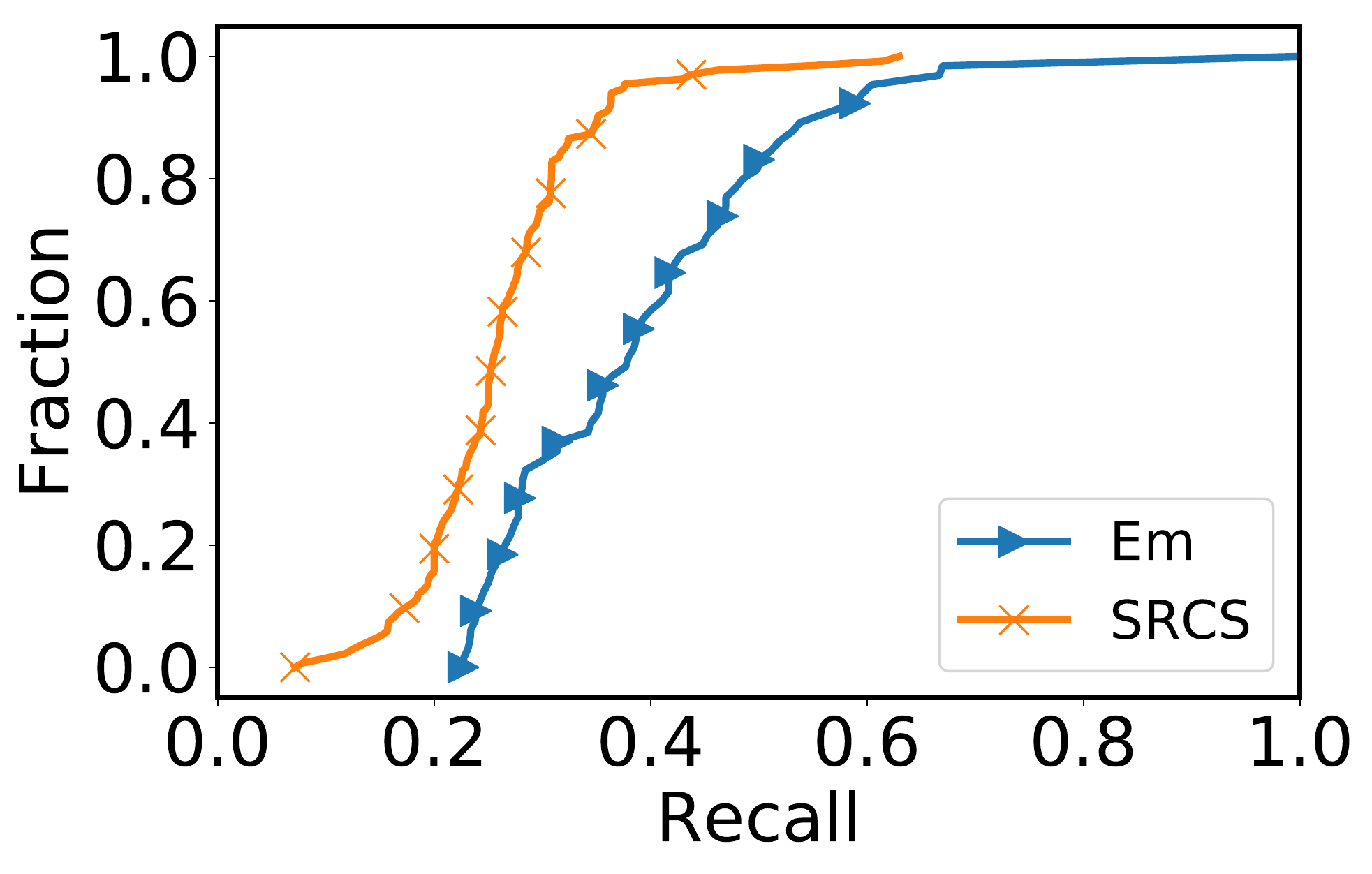}}
	\subfloat[Wikipedia, 50\% sampling]
	{\includegraphics[width=0.25\linewidth]{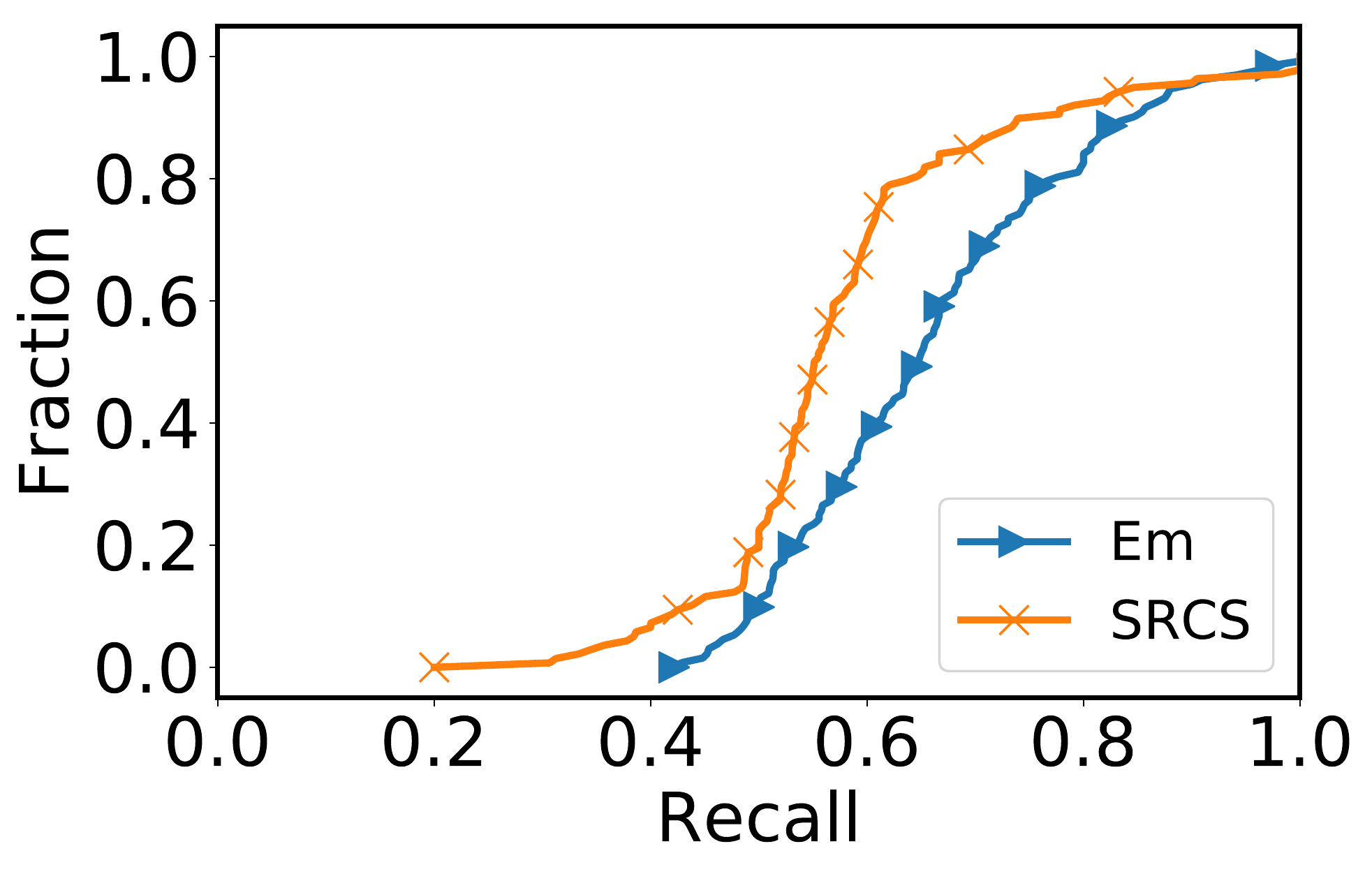}}
	\subfloat[CCNews, 25\% sampling]
	{\includegraphics[width=0.25\linewidth]{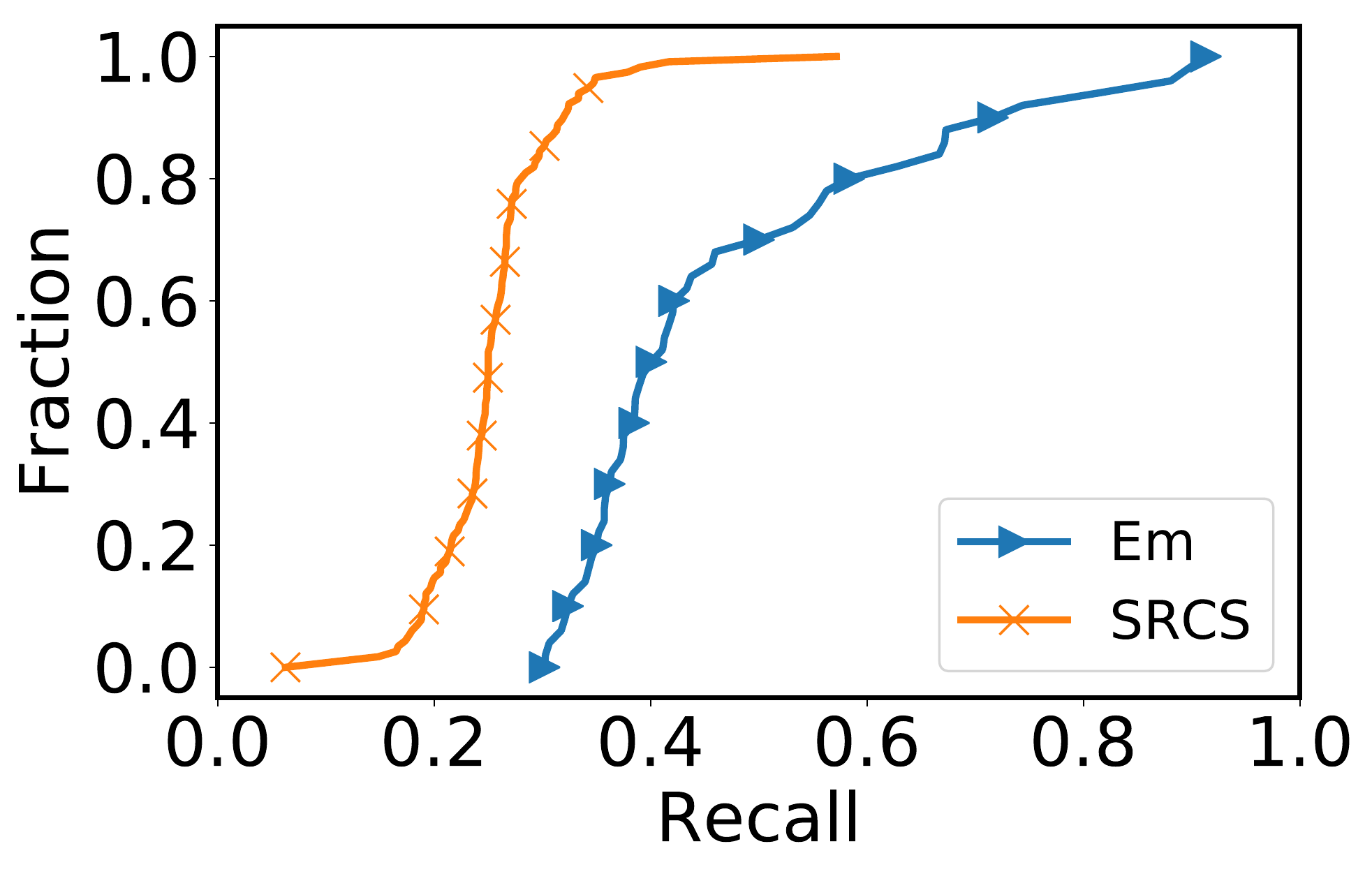}}
	\subfloat[CCNews, 50\% sampling]
	{\includegraphics[width=0.25\linewidth]{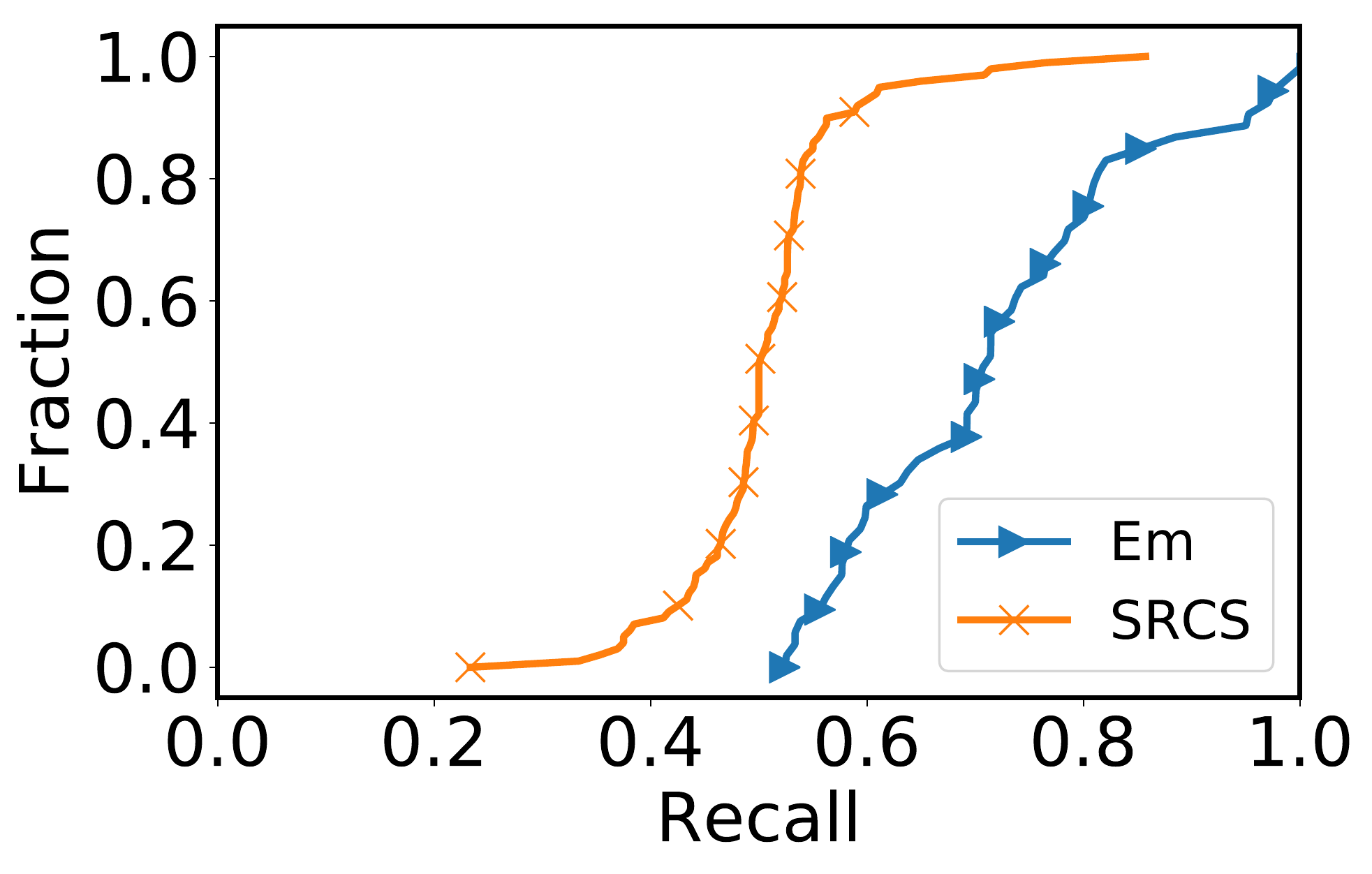}}
	\caption{CDFs of recall for Boolean retrieval queries over Wikipedia and CCNews data sets at different sampling rates.}
	\label{fig:BooleanCDF}
\end{figure*}

\begin{figure*}
	\subfloat[Avg. Speedup - Boolean] {\includegraphics[width=0.33\linewidth]{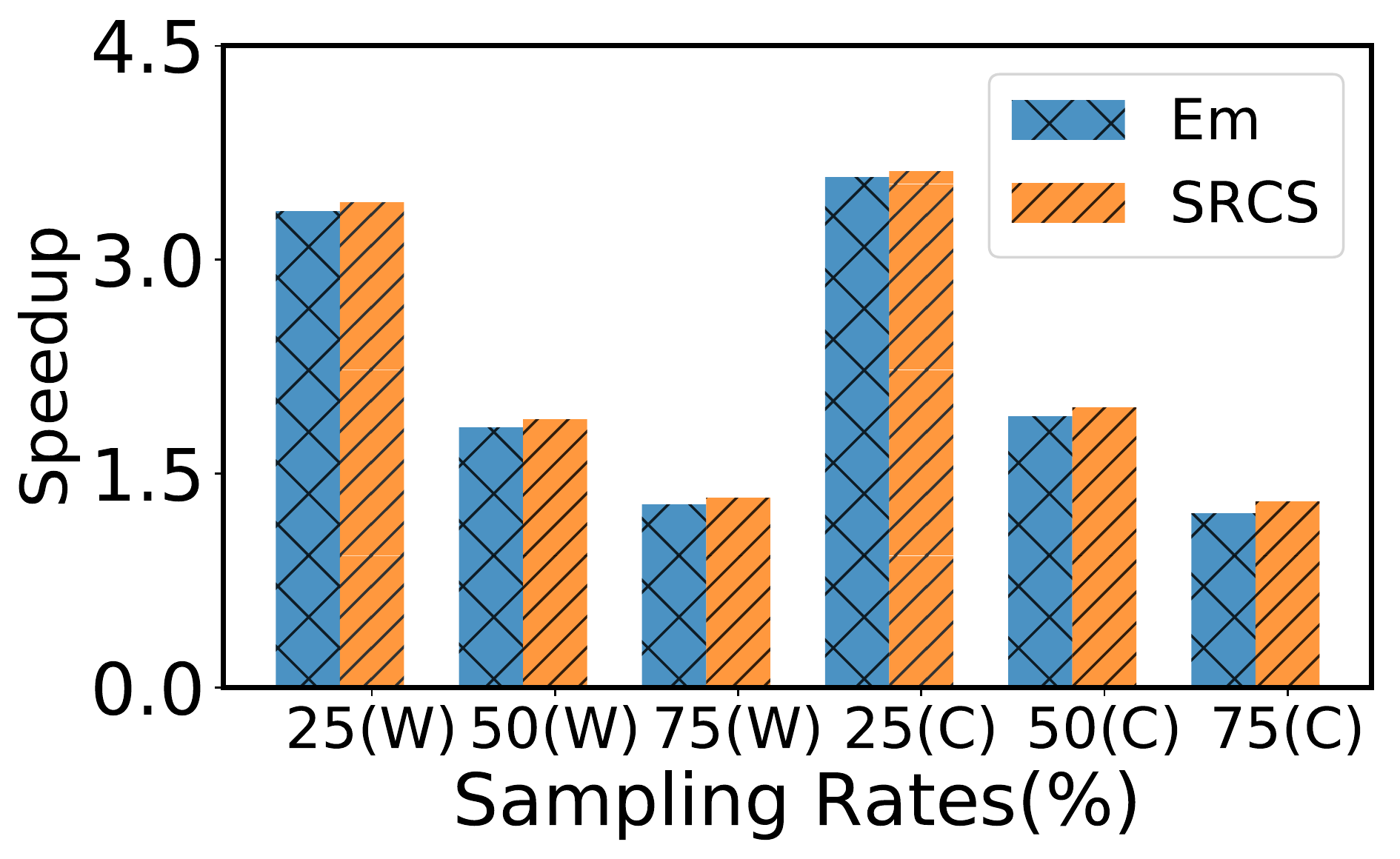}}
	\subfloat[Avg. Recall - Boolean]
	{\includegraphics[width=0.33\linewidth]{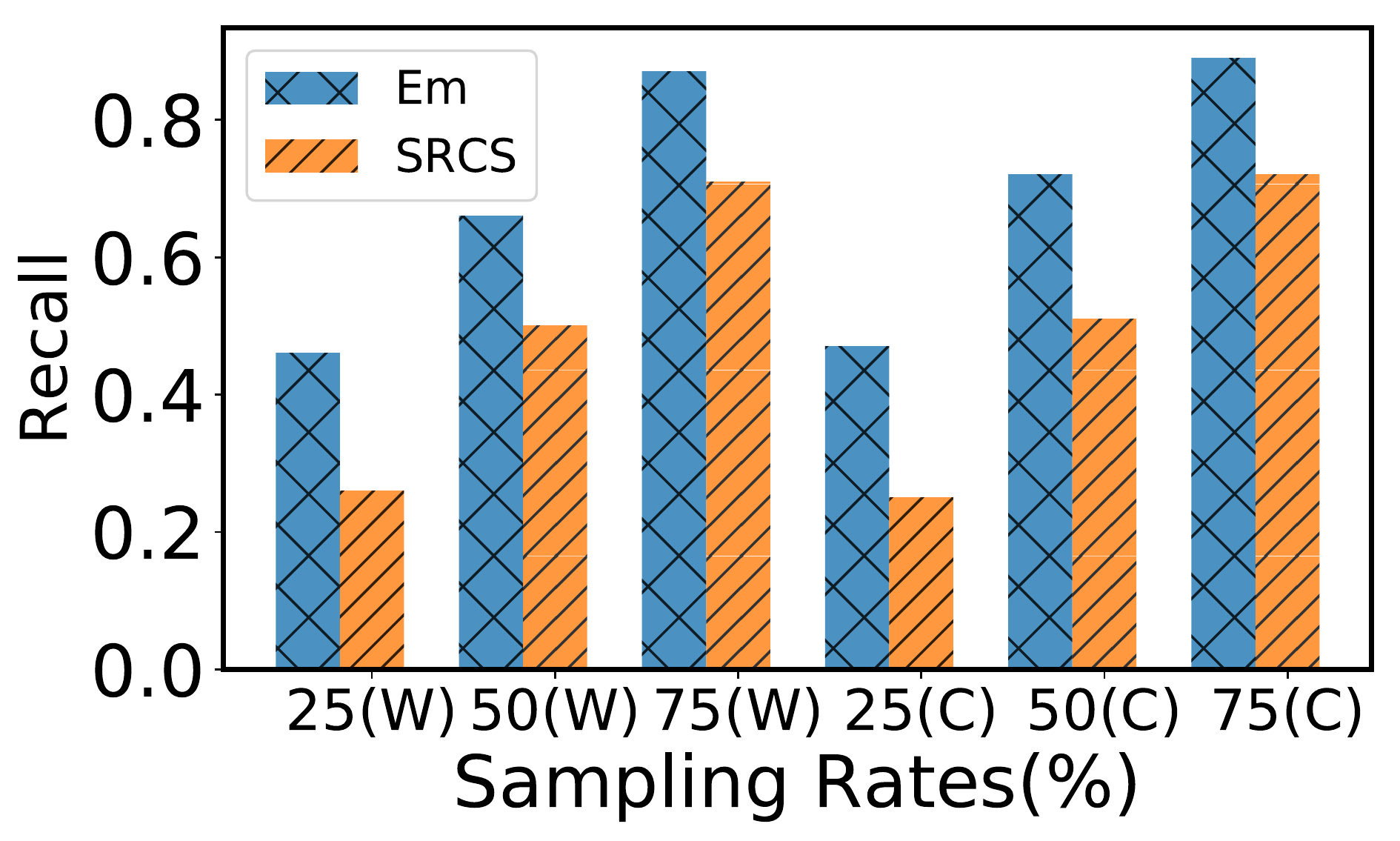}}
	\subfloat[Avg. P@10 - ranked] {\includegraphics[width=0.33\linewidth]{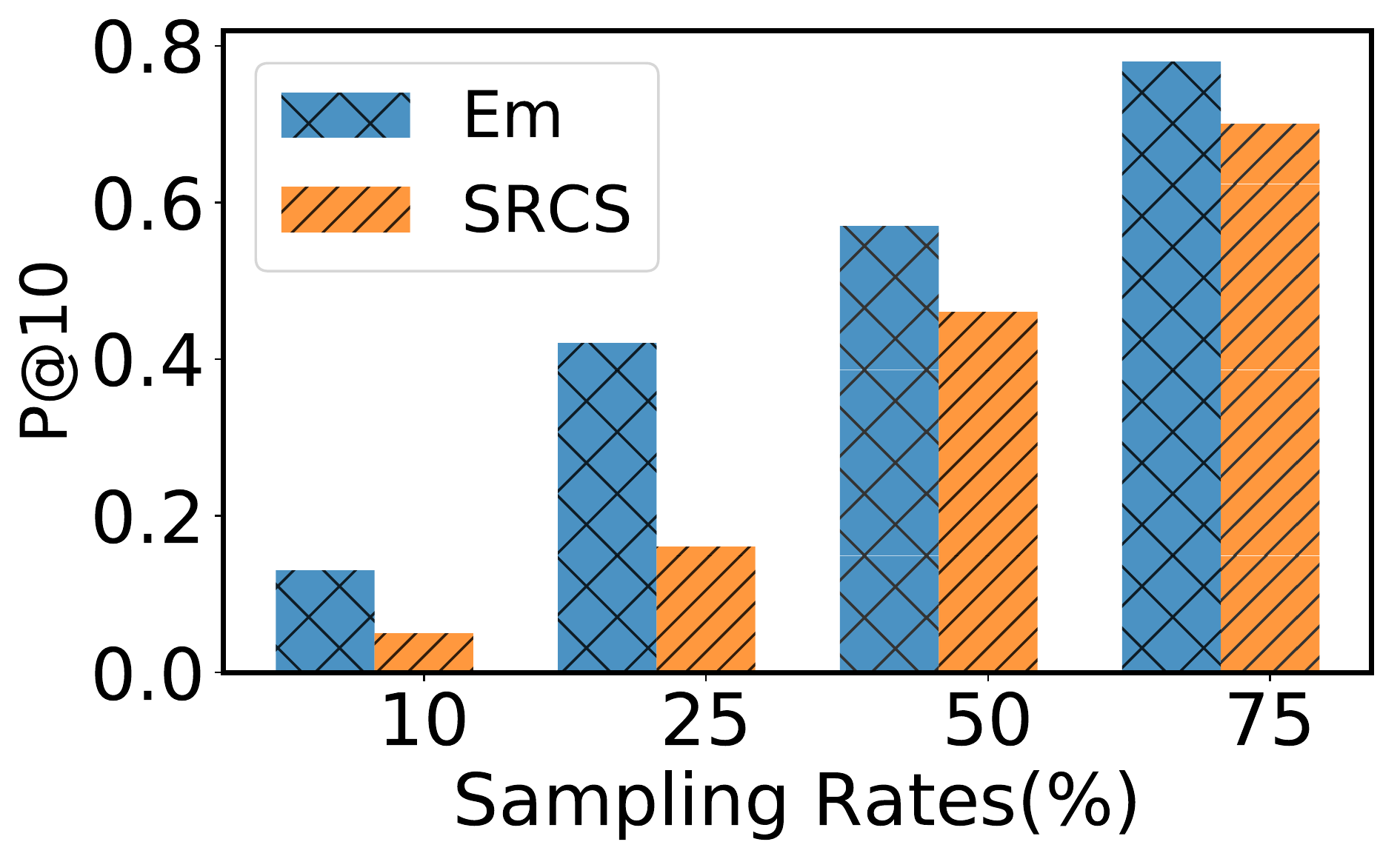}}
	\caption{(a) and (b) show speedup and recalls averaged across the test queries under 25\%, 50\% and 75\% sampling rates for Boolean retrieval~((W) and (C) represent Wikipedia and CCNews data sets respectively). (c) shows ranked retrieval precision P@10 achieved by EmApprox and SRCS at different sampling rates using Wikipedia data set.}
	\label{fig:IR}
\end{figure*}
\myparagraph{Recommendation queries.} We approximate user-centric CF, which takes as input a target user with past reviews/purchase history then outputs predicted ratings for unpurchased items. It also generates a top-k recommended item list sorted by their predicted ratings. \futurenote{\thunote{Review later whether this 1st paragraph is needed.}}

We rearrange the reviews in the original data set to group all reviews written by a unique user together. Each group of reviews written by a unique user is then considered a single document. We randomly select 100 users and remove 20\% of each selected user's ratings from the data set to be used as test data. We then cluster the remaining documents as we did for the DIR queries and construct the index.  Finally, we construct 100 queries for the selected users, where each query outputs the predicted ratings as computed by the CF algorithm for the users/reviews in the test data.

The rating scale is 1-5 in the Amazon data set. We report the mean squared error (MSE) and P@k to measure prediction performance. MSE is computed for the predicted vs.~actual ratings as a measure of accuracy for the predictions. P@k is the percentage of the items recommended in the top-k list that were actually purchased by the target user. (We assume that a user purchased an item if s/he reviewed it.)

\subsection{Results for aggregation queries}

Figure~\ref{fig:wcSamplingError1} plots the CDFs of estimated relative errors when running the 200 queries under EmApprox and SRCS at 1\%, 2.5\%, 5\% and 10\% sampling rates. We observe that: (1) EmApprox consistently achieves smaller estimated relative errors than SRCS at the same sampling rate; (2) the ``tails'' of the CDFs are ``shorter,'' meaning that there are fewer query answers with large estimated relative errors; and, (3) EmApprox achieves smaller maximum estimated relative errors. Under SRCS, the estimated relative errors can be large at very low sampling rates. For example, the estimated relative errors are 80\% and 95\% for the 50$^{th}$ (median) and 90$^{th}$ percentile, respectively, at 1\% sampling rate. Under EmApprox, they are reduced to 25\% and 45\%. Errors become much smaller with increasing sampling rates.

Figures~\ref{fig:wordCountAvgs}(a) and (b) show average speedups compared to precise execution and average relative errors (both estimated and actual), respectively. We observe that speedups are slightly smaller for EmApprox compared to SRCS. This is because EmApprox does incur a small amount of extra overhead to compute the sampling probabilities for blocks. On the other hand, as already mentioned, EmApprox achieves much smaller relative errors than SRCS. Specifically, SRCS has to process roughly 4x the amount of data processed by EmApprox to achieve similar relative errors: for example, the relative errors under EmApprox at 2.5\% sampling rate are similar to SRCS's relative errors at 10\% sampling rate.

In summary, EmApprox significantly outperforms SRCS. If the user can tolerate the estimated relative error profile for EmApprox at 10\% sampling rate (i.e., 30\% and 40\% at 50$^{th}$ and 90$^{th}$ percentiles, respectively), then EmApprox achieves an average speedup of $\sim$10x for the 200 aggregation queries. Further, the user can trade off between accuracy and performance by adjusting the sampling rate.

\subsection{Results for DIR queries}

Figure~\ref{fig:BooleanCDF} plots the CDFs of recall for the Boolean queries under EmApprox and SRCS when run on the Wikipedia (100 queries) and CCNews (100 queries) data sets. Similar to the observations for aggregation queries, we observe that EmApprox significantly outperforms SRCS. Unfortunately, even at a high sampling rate (e.g., 50\%), EmApprox misses significant fractions of relevant documents for many queries. This is because relevant documents can be spread out across many subcollections even with clustering, as clustering is performed based on overall contents of documents, which may be very different from the few words in a Boolean query.

Figures~\ref{fig:IR}(a) and (b) show Boolean retrieval's speedups over precise execution and the average achieved recall rates, respectively. We observe that the much higher sampling rates required to achieve higher recall rates constrain achievable speedups.

Figure~\ref{fig:IR}(c) shows the average P@10 results for ranked retrieval for the Wikipedia data set. EmApprox significantly outperforms SRCS at lower sampling rates, but even its achieved precision levels are unlikely to be acceptable. EmApprox achieves much better precision levels at higher sampling rates; e.g., 0.57 and 0.78 at 50\% and 75\% sampling rates, respectively.

In summary, EmApprox significantly outperforms SRCS. However, the nature of the problem, which is to find specific items in a large data set, reduces the effectiveness of sampling, even when sampling is directed by some knowledge of content. Thus, EmApprox can only achieve modest speedups while achieving relatively high recall rates and precision levels. For example, EmApprox achieves an average speedup of 1.3x at a sampling rate of 75\%. Achieved average recall and P@10 are 0.89 and 0.78, respectively.

\subsection{Results for recommendation queries}
\begin{figure}[!tbp]
	\subfloat[Avg. MSE]
	{\includegraphics[width=0.5\linewidth]{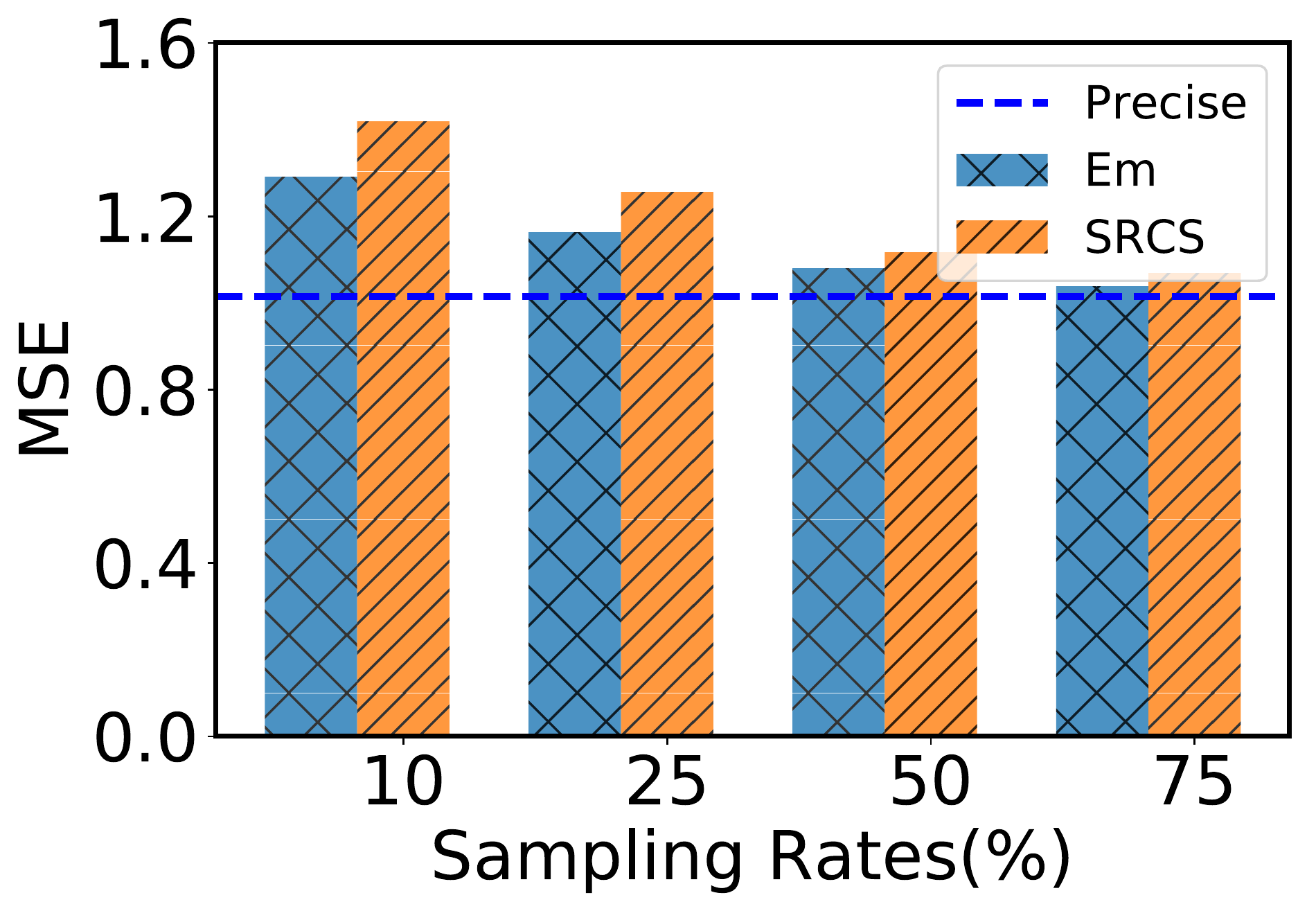}}
	\subfloat[Avg. P@10 (\%)]
	{\includegraphics[width=0.5\linewidth]{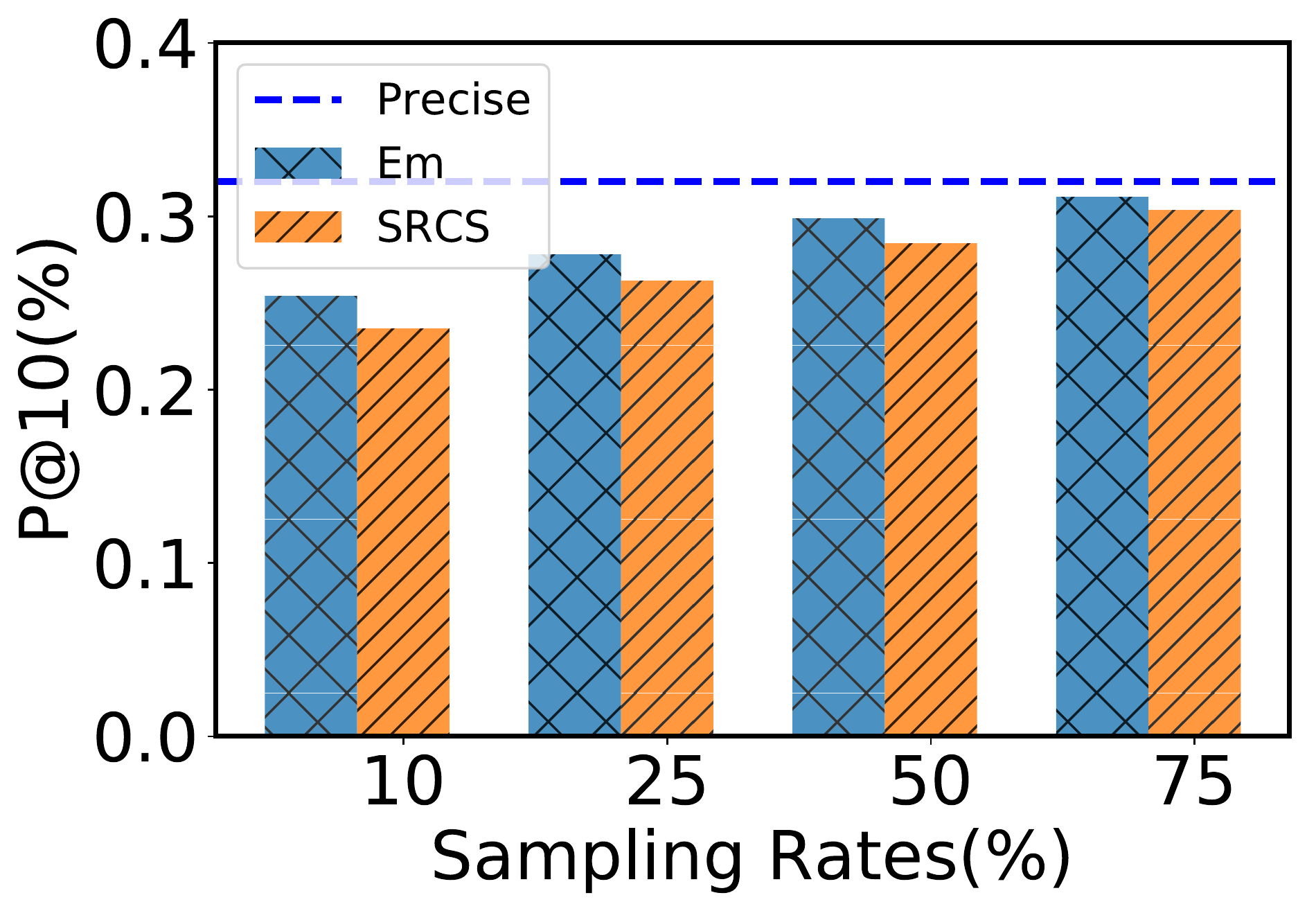}}
	\caption{Recommendation results showing avgerage prediction across the customers under different sampling rates including MSE and precision@10, where the horizontal line indicates the results under precise execution.}
	\label{fig:approxCF}
\end{figure}

Figure~\ref{fig:approxCF} shows average MSE and P@10 under different sampling rates for EmApprox and SRCS. Similar to results for the other two query types, EmApprox outperforms SRCS. The differences between the two approaches are less pronounced, however. For example, EmApprox outperforms SRCS by 8\% and 7.4\% for average MSE and average P@10, respectively, at a sampling rate of 25\%. This is likely due to the fact that user-centric CF itself does not achieve high accuracy---the precise execution achieves an average MSE of 1.015 and average P@10 of 0.32\%---so that selecting customers most similar to the target customer does not have a large impact compared to random selection. EmApprox incurs minimal overheads at query processing time, however, and so its increased accuracy is still desirable, especially when the number of customers is large. EmApprox speeds up the query processing time by almost 9x while degrading P@10 by 18.7\% at 10\% sampling rate. Speedup is over 3x with 12.5\% degradation of P@10 at 25\% sampling.

\begin{figure*}
	\vspace{-.5cm}
	\subfloat [Aggegation - Error]
	{\includegraphics[width = 0.25 \linewidth]{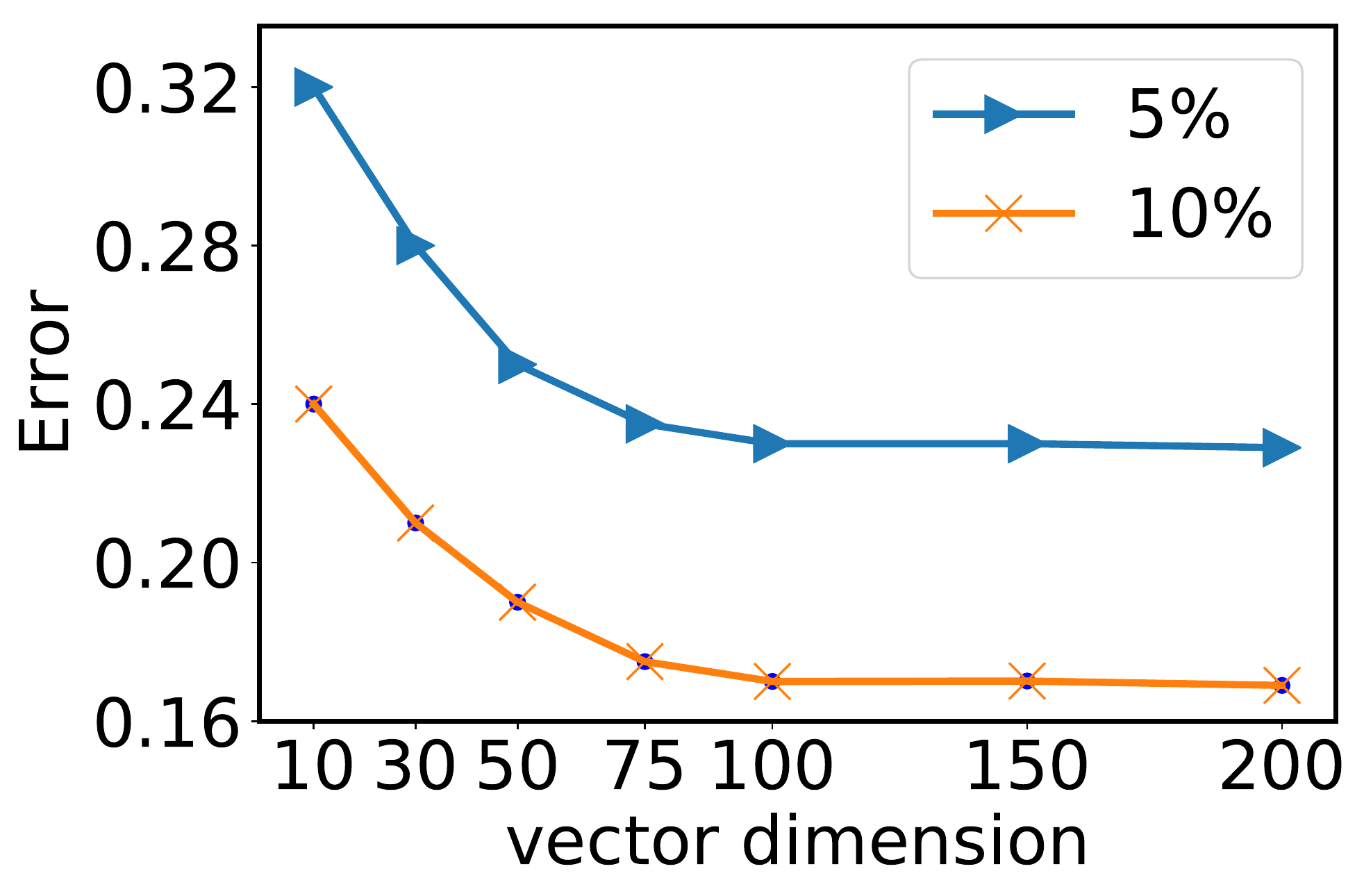}}
	\subfloat[Recommendation-MSE]
	{\includegraphics[width = 0.25 \linewidth]{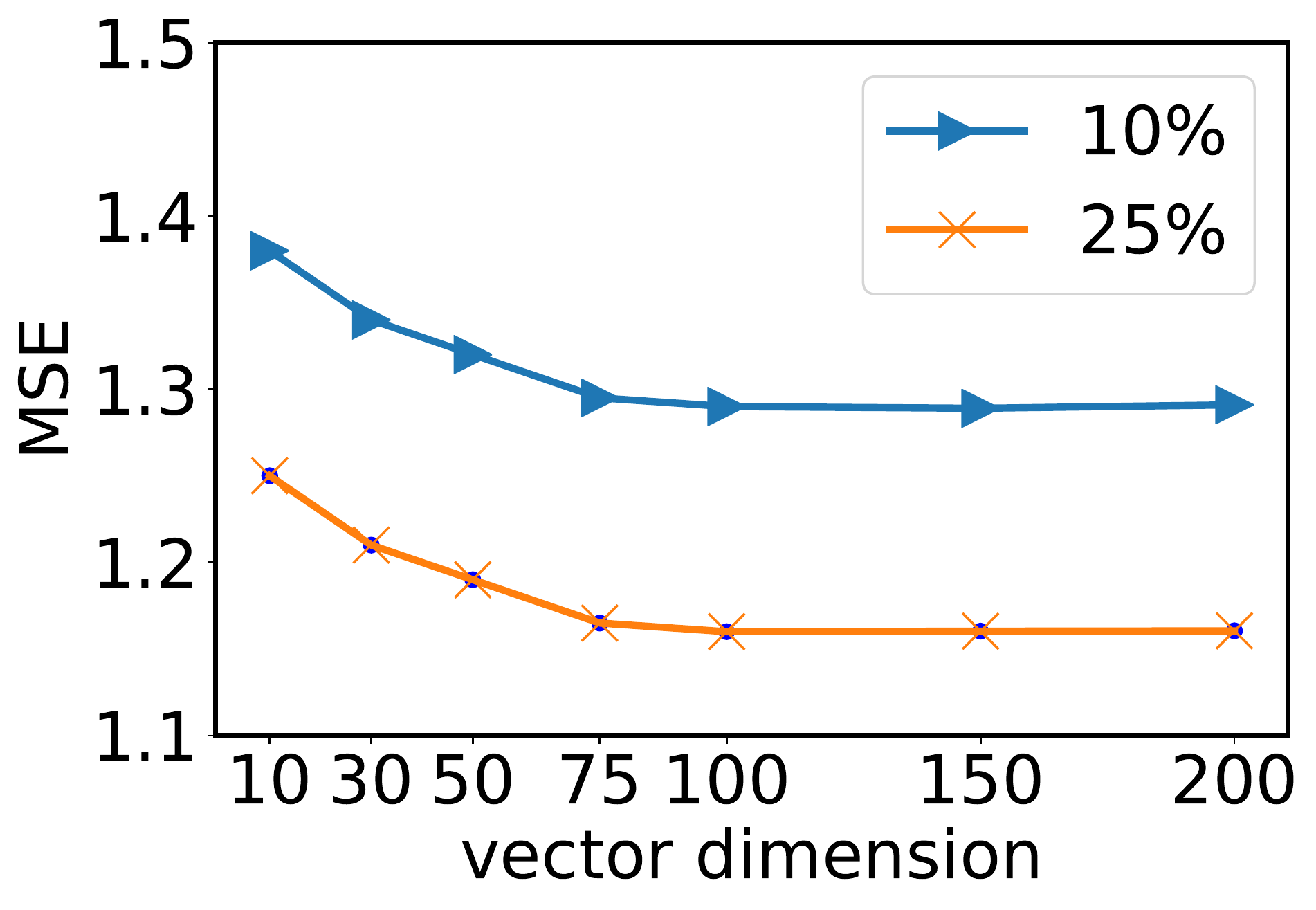}}
		\subfloat [Aggegation - Error]
	{\includegraphics[width = 0.25 \linewidth]{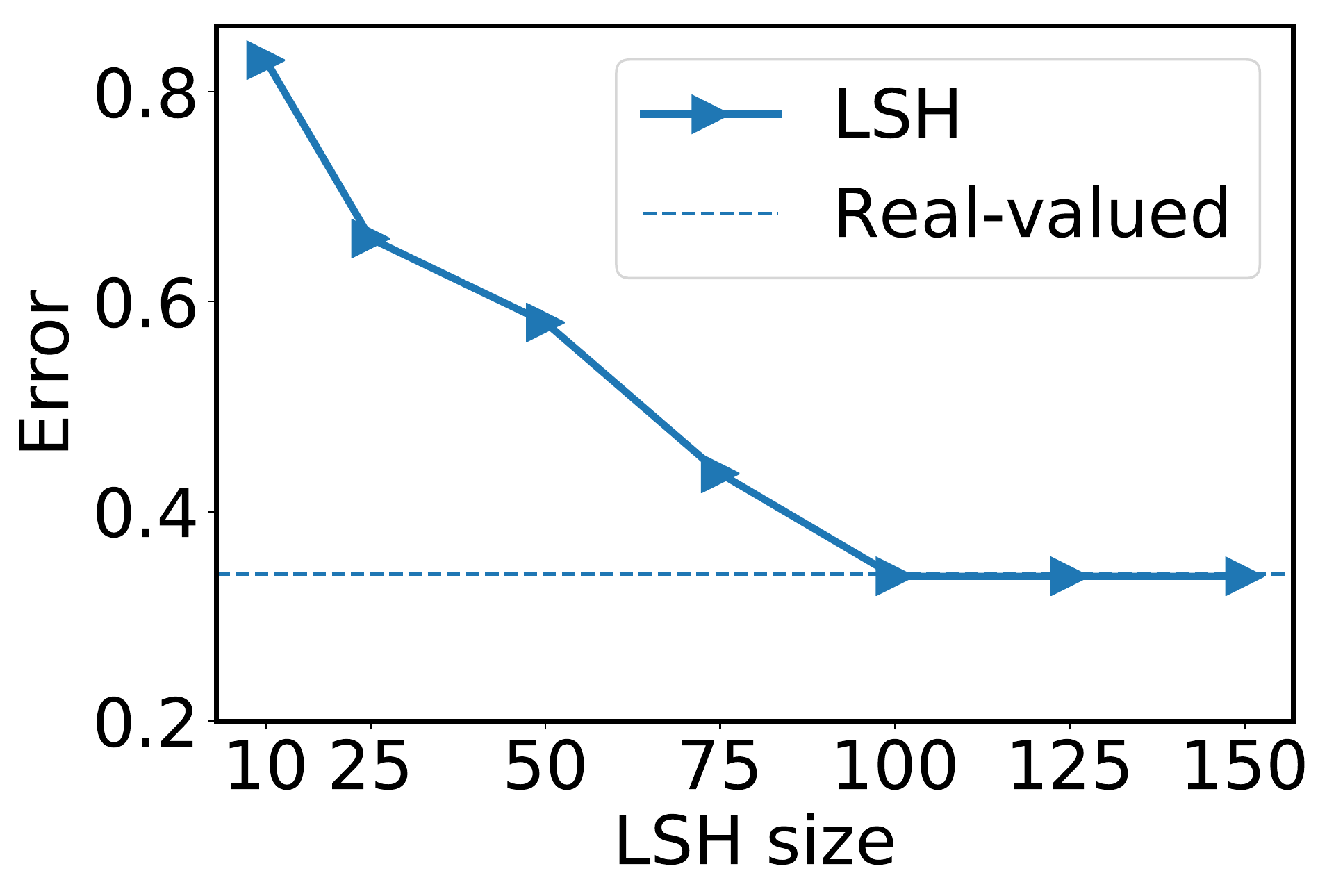}}
	\subfloat[Ranked retrieval - P@10]
	{\includegraphics[width = 0.25 \linewidth]{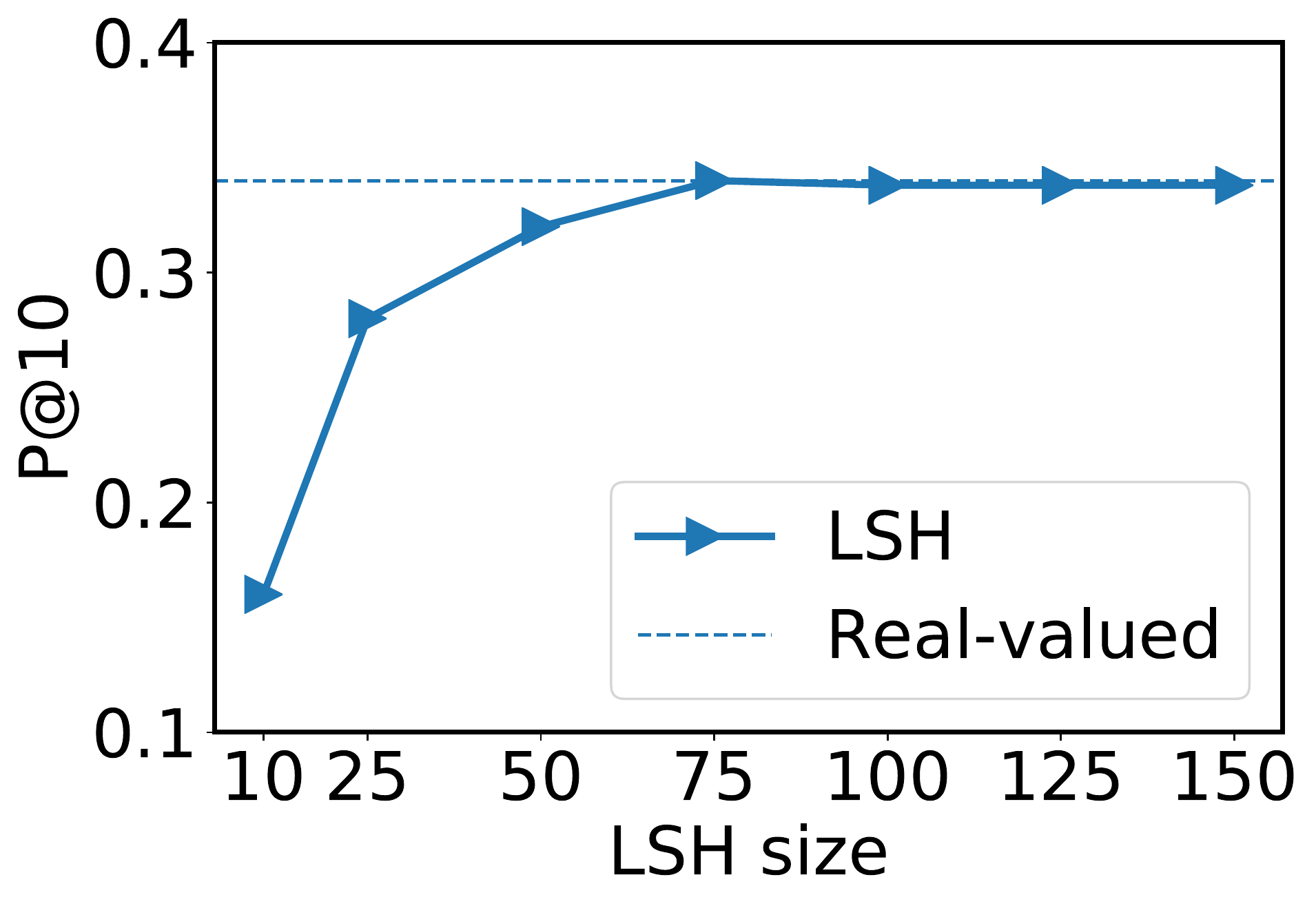}}
	\caption{(a) and (b) show the impact of vector dimension over the phrase occurrence~(aggregation) estimated error under 5\% and 10\% sampling rates, and MSE in the recommendation results under 10\% and 25\% sampling rates. (c) and (d) show the impact of LSH bits to the results using the Wikipedia data set under 1\% and 10\% sampling rates}
	\label{fig:sensitivity1}
\end{figure*}

\begin{figure}[!tbp]
	\subfloat [Ranked retrieval]
	{\includegraphics[width = 0.5 \linewidth]{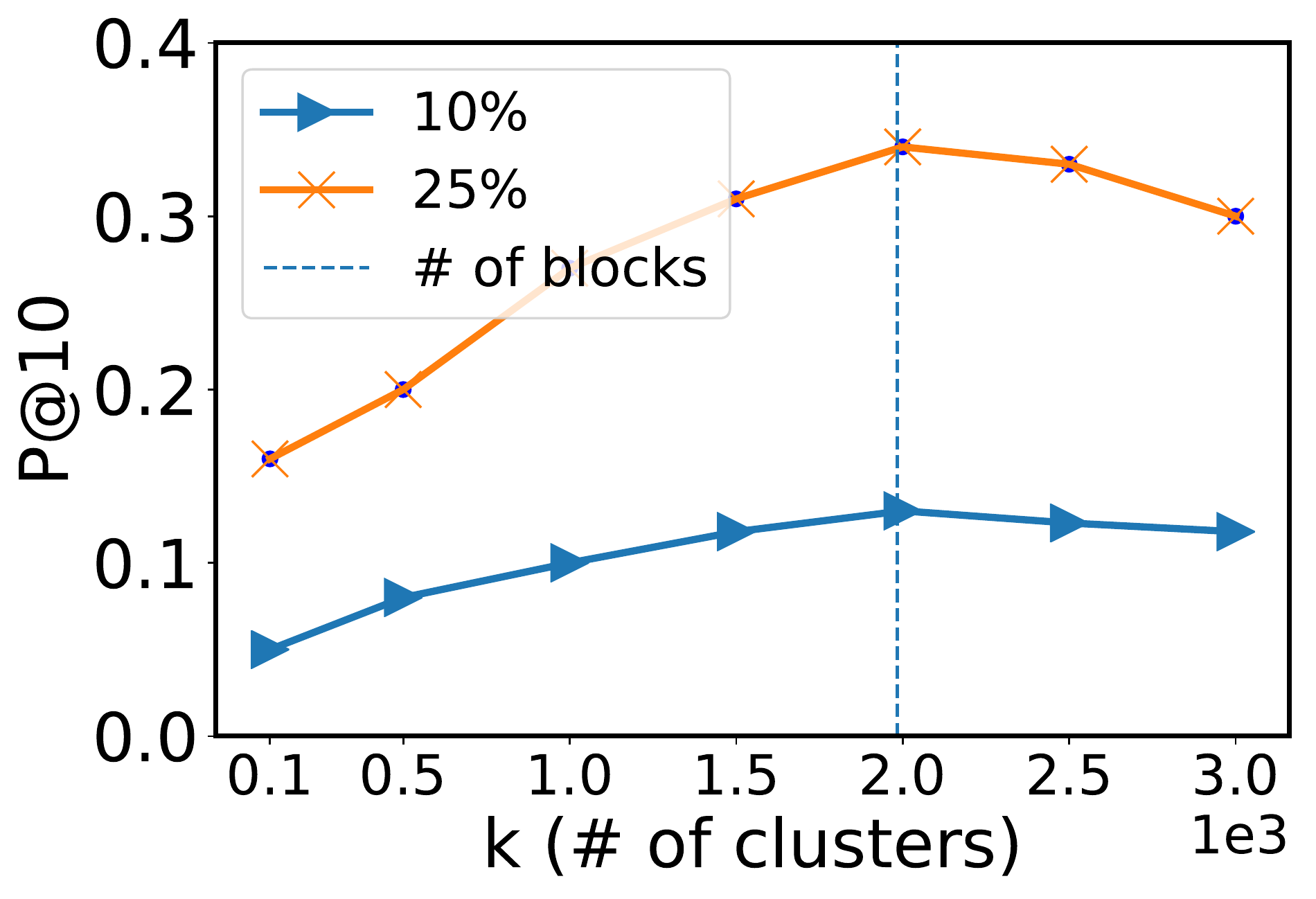}}
	\subfloat[Recommendation]
	{\includegraphics[width = 0.5 \linewidth]{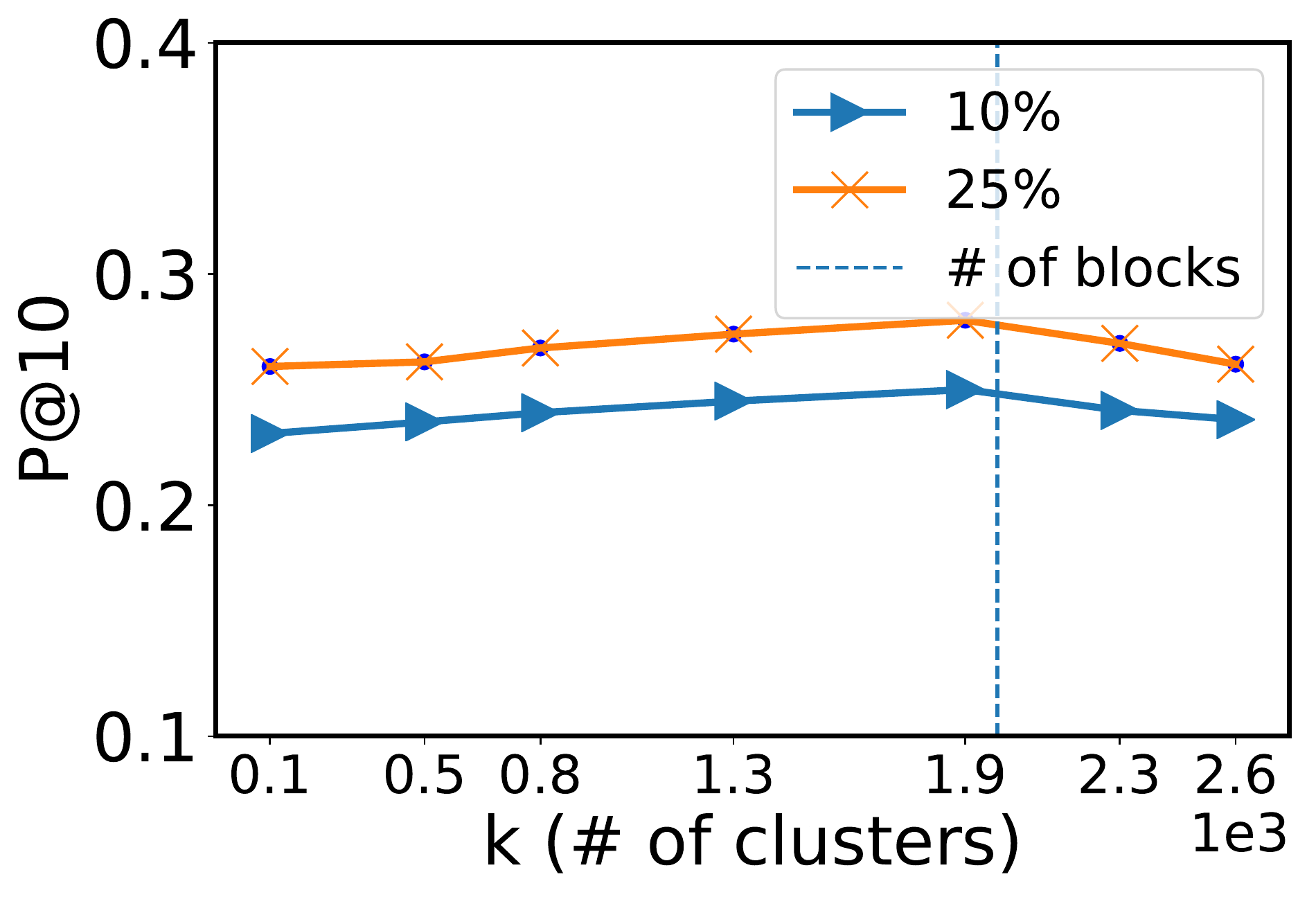}}
	
	\caption{Impact of k in the K-means clustering over P@10 for ranked retrieval~(Wikipedia) and CF, under 5\% and 25\% sampling rates.}
	\vspace{-0.5cm}
	\label{fig:sensitivity2}
\end{figure}

\subsection{Sensitivity Analysis}~\label{sec:sensitivity}
In this subsection, we study the effect of PV-DBOW and LSH bit vectors' dimensions~($\lambda_1$ and $\lambda_2$ respectively). For DIR and recommendation, we also study the number of clusters~($k$) in  K-means performed in offline preprocessing.

\myparagraph{Vector dimensions.}~Intuitively, $\lambda_1$ is more important than $\lambda_2$ as it directly reflects how accurate vectors can capture the word distributions in the data set. It is because mapping to LSH bit vectors is an approximation to its corresponding real-valued vectors thus is unlikely to enhance their performance. As a result, we first tune a suitable value of $\lambda_1$ and then $\lambda_2$. We find that the performance of EmApprox gradually improves with the increase of $\lambda_1$, and then stabilizes when its size is sufficiently large across our data sets. We would like a value of $\lambda_1$ that is as small as possible since large values will significantly slow down the training process of PV-DBOW. We choose 100 as the default vector size during training for convenience across the evaluations, since we observe that the performance of the queries are relatively stable when the vector size reaches 100. Note that the best value of $\lambda_1$ is still dependent on different data sets and queries combination.

Figures~\ref{fig:sensitivity1}(a) and (b) show the estimated error of phrase occurrence and MSE of the recommendation's relationship to the dimensions of vector. The observation for the effect of $\lambda_1$ is similar in tuning $\lambda_2$, where the performance first gradually increases then stabilizes as $\lambda_2$ gets larger. Our objective of tuning $\lambda_2$ is to match its performance with directly using real-valued vectors. Figure~\ref{fig:sensitivity1}(c) and (d) show the impact of $\lambda_2$ to the phrase occurrence and ranked retrieval queries. We notice that the smallest value of $\lambda_2$ to match the performance of its real-valued counterpart is related to both the data set and specific queries. For the experiments, we picked 100 for convenience because we have observed that this dimension has stably performed on par with the original real-valued vectors across the queries.

\myparagraph{Number of Clusters for K-means.}~We have found that the performance of EmApprox on DIR and recommendation queries is sensitive to $k$~(number of clusters) in K-means.  Figure~\ref{fig:sensitivity2}(a) and (b) show the P@10 metric for ranked retrieval over the Wikipedia data set and recommendation under different values of $k$. We observe that the performance gradually improves as $k$ increases and plateaus as it approaches the number of HDFS blocks in a data set. It is because when there are too many clusters, the data set will be close to have been randomly shuffled.

\subsection{Discussion}

Our results show that EmApprox is more effective for queries that estimate results from samples, than for queries that are seeking specific data items within large data sets. This matches intuition since EmApprox's index is meant to guide sampling toward subcollections that are most similar to a query, rather than identifying data items that are guaranteed to be relevant to the query as the traditional inverted index is meant to do. Thus, EmApprox can speed up aggregation and recommendation queries by up to 10x if the user can tolerate modest imprecision. In addition, the user can gracefully trade off imprecision with performance by adjusting the sampling rate. While we can observe the same trade offs for DIR, it is likely that users cannot tolerate the imprecision at lower sampling rates. For example, even at a sampling rate of 50\%, EmApprox achieves an average P@10 of less than 0.8, implying that the approximate processing misses over 4 of the top 10 documents. In comparison, under 10\% sampling, EmApprox achieves an estimated relative error of 18.2\% for aggregation, and a degradation on average of 18.7\% in P@10 for recommendation queries.

%% file: conclusion.tex
\section{Conclusion and future work}
We present an approximation framework for a wide range of analytical queries over large text data sets, using a light-weight index based on an NLP model~(PV-DBOW). We formally show that the training objective of PV-DBOW maximizes the generative probability of a query given a collection of documents. Our experiment shows that our light-weight index can reduce the execution time by almost an order of magnitude while degrading gracefully in approximation quality with decreasing sampling rates. EmApprox is particularly useful for exploratory text analytics.

\myparagraph{Future Work.}~We believe our proposed index can be effective under more scenarios and data sets. For example, our system can be used to efficiently detect the sentiment for a large text data set; it can also potentially be used toward selecting relevant users/products for more complex recommendation methods, such as model-based algorithms involving multiple data sources~\cite{McAuley:2013}. We also plan to extend our key methodology---learning an index directly from the data set to facilitate approximate computation---to other types of data sets, such as visual data, time series, etc.  

\reconsider{
In our work, we choose subcollections of the documents as sampling units, for significant reduction in execution time spent such as in I/O. LSH bit vectors can significantly reduce the storage space of the PV-DBOW model, and much faster to compute similarities than real-valued vectors, it makes it possible to sample at the document level, and the sampled documents may form new RDD partitions before a Spark job launches. This way the sampling decision is much more fine-grained and controlled, without the need to cluster the documents beforehand. We plan to explore this in our future work. We assume the dataset is stable, in the future we plan to handle updates to the dataset, where the learned models need to be incrementally trained. }

%% file: word2vec-based-approx.bbl
\begin{thebibliography}{10}
\providecommand{\url}[1]{#1}
\csname url@samestyle\endcsname
\providecommand{\newblock}{\relax}
\providecommand{\bibinfo}[2]{#2}
\providecommand{\BIBentrySTDinterwordspacing}{\spaceskip=0pt\relax}
\providecommand{\BIBentryALTinterwordstretchfactor}{4}
\providecommand{\BIBentryALTinterwordspacing}{\spaceskip=\fontdimen2\font plus
\BIBentryALTinterwordstretchfactor\fontdimen3\font minus
  \fontdimen4\font\relax}
\providecommand{\BIBforeignlanguage}[2]{{%
\expandafter\ifx\csname l@#1\endcsname\relax
\typeout{** WARNING: IEEEtran.bst: No hyphenation pattern has been}%
\typeout{** loaded for the language `#1'. Using the pattern for}%
\typeout{** the default language instead.}%
\else
\language=\csname l@#1\endcsname
\fi
#2}}
\providecommand{\BIBdecl}{\relax}
\BIBdecl

\bibitem{GNgram}
``{Google Book Ngrams},'' 2019,
  \url{http://storage.googleapis.com/books/ngrams/books/datasetsv2.html}.

\bibitem{CommonCrawl}
``Common crawl,'' \url{https://registry.opendata.aws/commoncrawl/}, 2018.

\bibitem{McAuley:2013}
\BIBentryALTinterwordspacing
J.~McAuley and J.~Leskovec, ``Hidden factors and hidden topics: Understanding
  rating dimensions with review text,'' in \emph{Proceedings of the 7th ACM
  Conference on Recommender Systems}, ser. RecSys '13.\hskip 1em plus 0.5em
  minus 0.4em\relax New York, NY, USA: ACM, 2013, pp. 165--172. [Online].
  Available: \url{http://doi.acm.org/10.1145/2507157.2507163}
\BIBentrySTDinterwordspacing

\bibitem{BlinkDB}
\BIBentryALTinterwordspacing
S.~Agarwal, B.~Mozafari, A.~Panda, H.~Milner, S.~Madden, and I.~Stoica,
  ``Blinkdb: Queries with bounded errors and bounded response times on very
  large data,'' in \emph{Proceedings of the 8th ACM European Conference on
  Computer Systems}, ser. EuroSys '13.\hskip 1em plus 0.5em minus 0.4em\relax
  New York, NY, USA: ACM, 2013, pp. 29--42. [Online]. Available:
  \url{http://doi.acm.org/10.1145/2465351.2465355}
\BIBentrySTDinterwordspacing

\bibitem{Sapprox}
\BIBentryALTinterwordspacing
X.~Zhang, J.~Wang, and J.~Yin, ``Sapprox: Enabling efficient and accurate
  approximations on sub-datasets with distribution-aware online sampling,''
  \emph{Proc. VLDB Endow.}, vol.~10, no.~3, pp. 109--120, Nov. 2016. [Online].
  Available: \url{https://doi.org/10.14778/3021924.3021928}
\BIBentrySTDinterwordspacing

\bibitem{ApproxHadoop}
\BIBentryALTinterwordspacing
I.~Goiri, R.~Bianchini, S.~Nagarakatte, and T.~D. Nguyen, ``Approxhadoop:
  Bringing approximations to mapreduce frameworks,'' in \emph{Proceedings of
  the Twentieth International Conference on Architectural Support for
  Programming Languages and Operating Systems}, ser. ASPLOS '15.\hskip 1em plus
  0.5em minus 0.4em\relax New York, NY, USA: ACM, 2015. [Online]. Available:
  \url{http://doi.acm.org/10.1145/2694344.2694351}
\BIBentrySTDinterwordspacing

\bibitem{peng2018aqp++}
J.~Peng, D.~Zhang, J.~Wang, and J.~Pei, ``Aqp++: connecting approximate query
  processing with aggregate precomputation for interactive analytics,'' in
  \emph{Proceedings of the 2018 International Conference on Management of
  Data}.\hskip 1em plus 0.5em minus 0.4em\relax ACM, 2018, pp. 1477--1492.

\bibitem{doc2vec}
Q.~Le and T.~Mikolov, ``Distributed representations of sentences and
  documents,'' in \emph{International Conference on Machine Learning}, 2014,
  pp. 1188--1196.

\bibitem{hdfs}
K.~Shvachko, H.~Kuang, S.~Radia, R.~Chansler \emph{et~al.}, ``The hadoop
  distributed file system.'' in \emph{MSST}, vol.~10, 2010, pp. 1--10.

\bibitem{ANNLsh}
\BIBentryALTinterwordspacing
P.~Indyk and R.~Motwani, ``Approximate nearest neighbors: Towards removing the
  curse of dimensionality,'' in \emph{Proceedings of the Thirtieth Annual ACM
  Symposium on Theory of Computing}, ser. STOC '98.\hskip 1em plus 0.5em minus
  0.4em\relax New York, NY, USA: ACM, 1998, pp. 604--613. [Online]. Available:
  \url{http://doi.acm.org/10.1145/276698.276876}
\BIBentrySTDinterwordspacing

\bibitem{ricci2015recommender}
F.~Ricci, L.~Rokach, and B.~Shapira, ``Recommender systems: introduction and
  challenges,'' in \emph{Recommender systems handbook}.\hskip 1em plus 0.5em
  minus 0.4em\relax Springer, 2015, pp. 1--34.

\bibitem{lohr2009sampling}
\BIBentryALTinterwordspacing
S.~Lohr, \emph{Sampling: Design and Analysis}, ser. Advanced (Cengage
  Learning).\hskip 1em plus 0.5em minus 0.4em\relax Cengage Learning, 2009.
  [Online]. Available: \url{https://books.google.com/books?id=aSXKXbyNlMQC}
\BIBentrySTDinterwordspacing

\bibitem{LearnedDBIndex}
T.~Kraska, A.~Beutel, E.~H. Chi, J.~Dean, and N.~Polyzotis, ``The case for
  learned index structures,'' in \emph{Proceedings of the 2018 International
  Conference on Management of Data}.\hskip 1em plus 0.5em minus 0.4em\relax
  ACM, 2018, pp. 489--504.

\bibitem{ApproxSpark}
G.~{Hu}, S.~{Rigo}, D.~{Zhang}, and T.~{Nguyen}, ``Approximation with error
  bounds in spark,'' in \emph{2019 IEEE 27th International Symposium on
  Modeling, Analysis, and Simulation of Computer and Telecommunication Systems
  (MASCOTS)}, Oct 2019, pp. 61--73.

\bibitem{word2vec}
T.~Mikolov, I.~Sutskever, K.~Chen, G.~S. Corrado, and J.~Dean, ``Distributed
  representations of words and phrases and their compositionality,'' in
  \emph{Advances in neural information processing systems}, 2013, pp.
  3111--3119.

\bibitem{paragraphVectorIR}
\BIBentryALTinterwordspacing
Q.~Ai, L.~Yang, J.~Guo, and W.~B. Croft, ``Analysis of the paragraph vector
  model for information retrieval,'' in \emph{Proceedings of the 2016 ACM
  International Conference on the Theory of Information Retrieval}, ser. ICTIR
  '16.\hskip 1em plus 0.5em minus 0.4em\relax New York, NY, USA: ACM, 2016, pp.
  133--142. [Online]. Available:
  \url{http://doi.acm.org/10.1145/2970398.2970409}
\BIBentrySTDinterwordspacing

\bibitem{NeuralMatrix}
\BIBentryALTinterwordspacing
O.~Levy and Y.~Goldberg, ``Neural word embedding as implicit matrix
  factorization,'' in \emph{Proceedings of the 27th International Conference on
  Neural Information Processing Systems - Volume 2}, ser. NIPS'14.\hskip 1em
  plus 0.5em minus 0.4em\relax Cambridge, MA, USA: MIT Press, 2014, pp.
  2177--2185. [Online]. Available:
  \url{http://dl.acm.org/citation.cfm?id=2969033.2969070}
\BIBentrySTDinterwordspacing

\bibitem{kulkarni2015selective}
A.~Kulkarni and J.~Callan, ``Selective search: Efficient and effective search
  of large textual collections,'' \emph{ACM Transactions on Information Systems
  (TOIS)}, vol.~33, no.~4, p.~17, 2015.

\bibitem{croft2015search}
W.~B. Croft, D.~Metzler, and T.~Strohman, \emph{Search engines: Information
  retrieval in practice}.\hskip 1em plus 0.5em minus 0.4em\relax Addison-Wesley
  Reading, 2015, vol. 283.

\bibitem{bm25}
S.~Robertson, H.~Zaragoza \emph{et~al.}, ``The probabilistic relevance
  framework: Bm25 and beyond,'' \emph{Foundations and Trends{\textregistered}
  in Information Retrieval}, vol.~3, no.~4, pp. 333--389, 2009.

\bibitem{ning2015comprehensive}
X.~Ning, C.~Desrosiers, and G.~Karypis, ``A comprehensive survey of
  neighborhood-based recommendation methods,'' in \emph{Recommender systems
  handbook}.\hskip 1em plus 0.5em minus 0.4em\relax Springer, 2015, pp. 37--76.

\bibitem{zhong2005efficient}
S.~Zhong, ``Efficient online spherical k-means clustering,'' in
  \emph{Proceedings. 2005 IEEE International Joint Conference on Neural
  Networks, 2005.}, vol.~5.\hskip 1em plus 0.5em minus 0.4em\relax IEEE, 2005,
  pp. 3180--3185.

\bibitem{hirschhorn2007gm}
M.~D. Hirschhorn, ``The am-gm inequality,'' \emph{The Mathematical
  Intelligencer}, vol.~29, no.~4, pp. 7--7, 2007.

\bibitem{PySpark}
``{PySpark},'' 2017,
  \url{https://spark.apache.org/docs/latest/api/python/index.html}.

\bibitem{Gensim}
``{Gensim},'' 2018, \url{https://radimrehurek.com/gensim/}.

\bibitem{abadi2016tensorflow}
M.~Abadi, P.~Barham, J.~Chen, Z.~Chen, A.~Davis, J.~Dean, M.~Devin,
  S.~Ghemawat, G.~Irving, M.~Isard \emph{et~al.}, ``Tensorflow: A system for
  large-scale machine learning,'' in \emph{12th $\{$USENIX$\}$ Symposium on
  Operating Systems Design and Implementation ($\{$OSDI$\}$ 16)}, 2016, pp.
  265--283.

\bibitem{Wikipedia}
``Wikipedia database,'' \url{http://en.
  wikipedia.org/wiki/Wikipedia_database.}, 2018.

\bibitem{CCNews}
``Common crawl news dataset,''
  \url{http://commoncrawl.org/2016/10/news-dataset-available.}, 2016.

\bibitem{Solr}
``{Apache Solr},'' 2019, \url{http://lucene.apache.org/solr/}.

\bibitem{Lucene}
``{http://lucene.apache.org/},'' 2019, \url{http://lucene.apache.org/}.

\end{thebibliography}
